\documentclass[aps,pra,reprint,groupedaddress,amsmath]{revtex4-1}

\usepackage{graphicx}
\usepackage{dcolumn}
\usepackage{amsmath}
\usepackage{lipsum}

\begin{document}


\title{Characterization of a Simultaneous Dual-Species Atom Interferometer for a Quantum Test of the Weak Equivalence Principle}


\author{A. Bonnin}
\email{alexis.bonnin@onera.fr}
\affiliation{ONERA, DMPH, BP 80100, 91123, Palaiseau, France}

\author{N. Zahzam}
\affiliation{ONERA, DMPH, BP 80100, 91123, Palaiseau, France}

\author{Y. Bidel}
\affiliation{ONERA, DMPH, BP 80100, 91123, Palaiseau, France}

\author{A. Bresson}
\affiliation{ONERA, DMPH, BP 80100, 91123, Palaiseau, France}


\date{\today}

\begin{abstract}
We present here the performance of a simultaneous dual-species matter-wave accelerometer for measuring the differential acceleration between two different atomic species ($^{87}$Rb and $^{85}$Rb). We study the expression and the extraction of the differential phase from the interferometer output. The differential accelerometer reaches a short-term sensitivity of $1.23\times10^{-7}g/\sqrt{Hz}$ limited by the detection noise and a resolution of $2\times10^{-9}g$ after 11000 s, the highest reported thus far with a dual-species atom interferometer to our knowledge. Thanks to the simultaneous measurement, such resolution levels can still be achieved even with vibration levels up to $3\times10^{-3}g$, corresponding to a common-mode vibration noise rejection ratio of 94 dB (rejection factor of 50 000). These results prove the ability of such atom sensors for realizing a quantum based test of the weak equivalence principle (WEP) at a level of $\eta\sim10^{-9}$ even with high vibration levels and a compact sensor.
\end{abstract}


\maketitle



\section{Introduction}

Light pulse atom interferometry \cite{Borde1989,Tino2014} has allowed to probe, in an extremely sensitive and accurate way, the incidence of inertial forces on the motion of cold atoms. In recent decades, the very high performance of such sensors have been demonstrated for measuring the gravity acceleration \cite{Peters2001,Gillot2014,Hu2013,Bidel2013,Hauth2013}, Earth's gravity gradient \cite{McGuirk2002,Sorrentino2014,Duan2014}, rotations \cite{Gustavson1997,Gauguet2009,Tackmann2012} and gravity-field curvature \cite{Rosi2015}. That is why they have been greatly developed for many applications such as inertial navigation \cite{Jeleki2005}, geophysics, geodesy, sub-surface exploration, and metrology \cite{Gillot2014}.

Many experiments and projects worldwide, also prove that quantum sensors appear to be very promising tools for exploring many aspects of fundamental physics as the determination of the fine-structure constant $\alpha$ \cite{Bouchendira2011} and the Newtonian gravity constant $G$ \cite{Fixler2007,Rosi2014}, the detection of gravitational waves \cite{Dimopoulos2008}, the exploration of short-range forces \cite{Ferrari2006,Wolf2007} and quantum based tests of the Weak Equivalence Principle (WEP) \cite{Fray2004,Bonnin2013,Schlippert2014,Tarallo2014,Zhou2015}.

The WEP, also called the Universality of Free Fall (UFF), is one of the pillar of the Einstein's theory of general relativity. This postulate states that all bodies, regardless of their internal composition, are affected by gravity in a universal way, in particular all test particles at the alike space-time point in a given gravitational field will undergo the same acceleration. Today, different theories predict its violation \cite{Damour2012,Hohensee2013} in the purpose of unifying general relativity with non-gravitational fundamental interactions as described by the standard model in a quantum approach. Matter-wave tests thus offer new alternatives by their intrinsically quantum nature, radically different from their classical counterparts \cite{Schlippert2014}, that can bring new constraints and bounds on the unifying theories.

Presently, the state-of-the-art WEP test by atom interferometry is at a level of few $10^{-8}$ \cite{Zhou2015}. In order to improve the sensitivity of atom accelerometers, several projects under development aim to compare the free fall of two different atomic species during few seconds, as the sensitivity scales quadratically with the interrogation time, in 10-m-tall atomic fountains \cite{Dimopoulos2007,Zhou2011,Hartwig2015}, drop towers \cite{Muntinga2013}, sounding rockets, parabolic flights \cite{Geiger2011} and in space \cite{Tino2013,Altschul2015,Aguilera2014}. These projects are developed in parallel with works on the increase of the momentum transfer \cite{Leveque2009,Chiow2011,Clade2009,Muller2009} to enhance the enclosed area and further improve the interferometers sensitivities. The use of Bose Einstein condensates as atomic sources is also prospected \cite{Hardman2014,Kuhn2014,Muntinga2013}.

Today, the sensitivity of most state-of-the-art gravimeters \cite{Muller2008,Gillot2014} is limited by vibration noise. To avoid this effect in quantum based test of the WEP, a special attention must be paid to the simultaneous interrogation of both quantum proof bodies in order to benefit from an efficient common-mode noise rejection allowed by the differential measurement \cite{Bonnin2013}.

In this paper, we present the characterization of a simultaneous dual-species atom interferometer which simultaneously handles both isotopes of rubidium ($^{87}$Rb and $^{85}$Rb). This sensor allows us to study the specificities of a differential acceleration measurement, and especially the rejection of the vibration noise which is currently at 94 dB. We also show that the achieved sensitivity of our experiment is compatible with a test of the WEP at a level of $2\times10^{-9}$. This work is part of the context of quantum based tests of the WEP and contributes to the elaboration of future experiments, both ground and space, aiming to detect a WEP violation.

The paper is organized as following: section \ref{Differential phase expression in a dual-species atom interferometer} gives the formalism for calculating a complete expression of the differential phase in our dual-species Mach-Zehnder type interferometer. Section \ref{Differential phase extraction} explains then how this differential phase is extracted from the elliptic interferometric signal. Section \ref{Experimental Apparatus} describes the experimental apparatus and the overall measurement sequence. Section \ref{Previous results and improvements} presents the improvements brought to the experiment leading to the new results on the differential acceleration measurement given in section \ref{Resolution, sensitivity and long term stability}. Finally, section \ref{Rejection of vibration noise} deals about the common-mode vibration noise rejection and presents its theoretical and experimental limits. A brief discussion on a quantum test of the WEP is given as a conclusion in section \ref{Conclusion}.

\section{Differential phase expression in a dual-species atom interferometer}
\label{Differential phase expression in a dual-species atom interferometer}

For simultaneously measuring the acceleration undergone by $^{87}$Rb and $^{85}$Rb, a Mach-Zehnder type atom interferometer is realized. This interferometer consists of a sequence of three equally spaced light pulses, of duration $\tau-2\tau-\tau$, driving stimulated Raman transitions between the two fundamental hyperfine states of the atoms \cite{Kasevich1992}. At the end, considering the isotope $i$, the proportion of atoms in each state depends sinusoidally on the phase difference between both paths of the interferometer $\Delta \Phi_{i}$, $\Delta \Phi_{i}$ being proportional to the acceleration $a_{i}$ undergone by the isotope $i$ along the Raman laser direction of propagation. To first approximation, by considering infinitely short Raman pulses (\textit{i.e.} $\tau \ll T$), the phase is given by \cite{Peters2001}
\begin{equation}\label{interferometer phase}
\Delta \Phi_{i} \simeq (\vec{k}^{i}_{\scriptstyle\textrm{eff}}\cdot\vec{a}_{i} - 2\pi \alpha)T^{2}
\end{equation}
where $\vec{k}^{i}_{\scriptstyle\textrm{eff}}$ is the effective wave vector of the Raman laser associated to isotope $i$, $\alpha$ is the microwave chirp applied on the Raman frequency to compensate for the Doppler shift induced by gravity for both isotopes, and $T$ is the time between two pulses of light.

The two signals from the dual-species atom interferometer are sinusoidal functions of the interferometric phase and can be expressed as
\begin{equation}\label{fringes expression}
\left\{
	\begin{array}{ll}
			\vspace{0cm}
			P_{87} = P_{87}^{0} - \frac{C_{87}}{2}\cos(\Delta\Phi_{87})
			\vspace{3pt} \\
			P_{85} = P_{85}^{0} - \frac{C_{85}}{2}\cos(\Delta\Phi_{87}+\phi_{d})
	\end{array}
\right.\
\end{equation}
with
\begin{equation}\label{differential phase}
\phi_{d}=\Delta\Phi_{85}-\Delta\Phi_{87}
\end{equation}
where $P_{87{,}85}$ are the proportions of atoms in the upper hyperfine ground state, $P^{0}_{87{,}85}$ are the offsets of the population measurements, $C_{87{,}85}$ are the fringe amplitudes, $\Delta\Phi_{87}$ is the interferometric phase for $^{87}$Rb as expressed in Eq.(\ref{interferometer phase}) and $\phi_{d}$ represents the differential phase between the two species. We are interested in measuring this last quantity which is directly linked to the differential acceleration undergone by atoms. 

The sensitivity function formalism \cite{Cheinet2008} has been used in this paper in order to express the more strictly as possible this differential phase. This formalism allows to take into account the finite duration of the Raman pulses. The sensitivity function $g_{s}(t)$ is linked to the step response of the atom interferometer to an infinitesimal variation $\delta\varphi$ of the Raman laser phase. In this paper, we are interested in the impulse response of the Mach-Zehnder type atom interferometer to accelerations. These responses are given by the response functions $f_{87,85}(t)$ associated to each isotopes which are basically equal to the primitive integrals of the previous sensitivity functions. The interferometric phases can thus be expressed by following this formalism,
\begin{equation}\label{phase with response functions}
\left\{
	\begin{array}{l}
	\vspace{0.1cm}
			\Delta\Phi_{87}=\int f_{87}(t)(\vec{k}_{\scriptstyle\textrm{eff}}^{87}\cdot\vec{a}_{87}(t) - 2\pi \alpha)\,dt+\phi^{87}_{SE}
			\\
			\Delta\Phi_{85}=\int f_{85}(t)(\vec{k}_{\scriptstyle\textrm{eff}}^{85}\cdot\vec{a}_{85}(t) - 2\pi \alpha)\,dt+\phi^{85}_{SE}
	\end{array}
\right.\
\end{equation}
with $a_{87}(t)=a+\tilde{a}(t)$ and $a_{85}(t)=a+\Delta a+\tilde{a}(t)$. $a_{87,85}(t)$ are the accelerations of each isotope projected along $\vec{k}_{\scriptstyle\textrm{eff}}^{87,85}$ relative to the inertial reference corresponding to the Raman mirror in our experiment. They are composed of terms of different origins: the mutual constant acceleration $a$ (corresponding to the gravity acceleration $g$ for a ground gravimeter and to 0 for a whole instrument in perfect free fall), the additional acceleration of the Raman mirror $\tilde{a}(t)$ and the potential WEP violation signal $\Delta a$ (as defined in Eq. (\ref{Eotvos ratio})). For its part, $\phi^{87,85}_{SE}$ represent the phase shifts due to systematic effects.

The calculation and the demonstration of the sensitivity function have already been carefully described in \cite{Cheinet2008}, we have slightly extended the experimental framework wherein this function is calculated. We assume here a case where the Rabi pulsations $\Omega_{1}$, $\Omega_{2}$, $\Omega_{3}$, associated to each of the three pulses of the interferometer, are not equal one to another ($\Omega_{1}\neq\Omega_{2}\neq\Omega_{3}$) and do not realize a perfect $\pi/2$-$\pi$-$\pi/2$ sequence ($\Omega_{1,2,3}\tau\neq\pi/2$). Experimentally, these differences in Rabi frequencies may come from many sources. For instance, they can be induced by Raman laser power fluctuations between each light pulse. The transverse motion of an atom may also be a source of Rabi pulsation differences because of the spatial inhomogeneity of the Raman laser beam intensity. In our experiment, these differences are mainly due to additional laser lines which generate a spatial dependency of the Rabi frequency. This point will be discussed more in details in section \ref{Previous results and improvements} and Appendix \ref{Appendix additional laser lines}. In this framework, the expression of the sensitivity function of the interferometer can be derived from the following definition :

\begin{equation}\label{sensitivity function definition}
g_{s}^{i}(t)=\frac{2}{C_{i}\sin\left(\Delta \Phi_{i}\right)} \lim\limits_{\substack{\delta\varphi\rightarrow0}} \frac{\delta P_{i}\left(\delta\varphi,t\right)}{\delta\varphi}=\lim\limits_{\substack{\delta\varphi\rightarrow0}} \frac{\Delta \Phi_{i}\left(\delta\varphi,t\right)}{\delta\varphi}
\end{equation}

From the sensitivity function, the interferometric phase $\Delta \Phi_{i}$ can be evaluated for arbitrary evolution of the phase of the Raman laser $\varphi(t)$ \cite{Cheinet2008}

\begin{equation}\label{sensitivity function and interferometric phase}
\Delta \Phi_{i}=\int g_{s}^{i}(t) \,d\varphi(t)=\int g_{s}^{i}(t)\frac{d\varphi(t)}{dt}\,dt
\end{equation}

In the frame associated to the free falling atom, the phase of the Raman laser experienced by the atom is given by $\varphi(t)=\vec{k}^{i}_{\scriptstyle\textrm{eff}}\cdot\vec{r}_{i}(t)$, where $\vec{r}_{i}(t)$ is the position of the atoms $i$ compare to the inertial reference. The equation (\ref{sensitivity function and interferometric phase}) thus shows that the sensitivity function corresponds to the impulse response of the interferometer relative to the velocity of the free falling atom.

\begin{figure}
\centerline{\includegraphics[width=8cm]{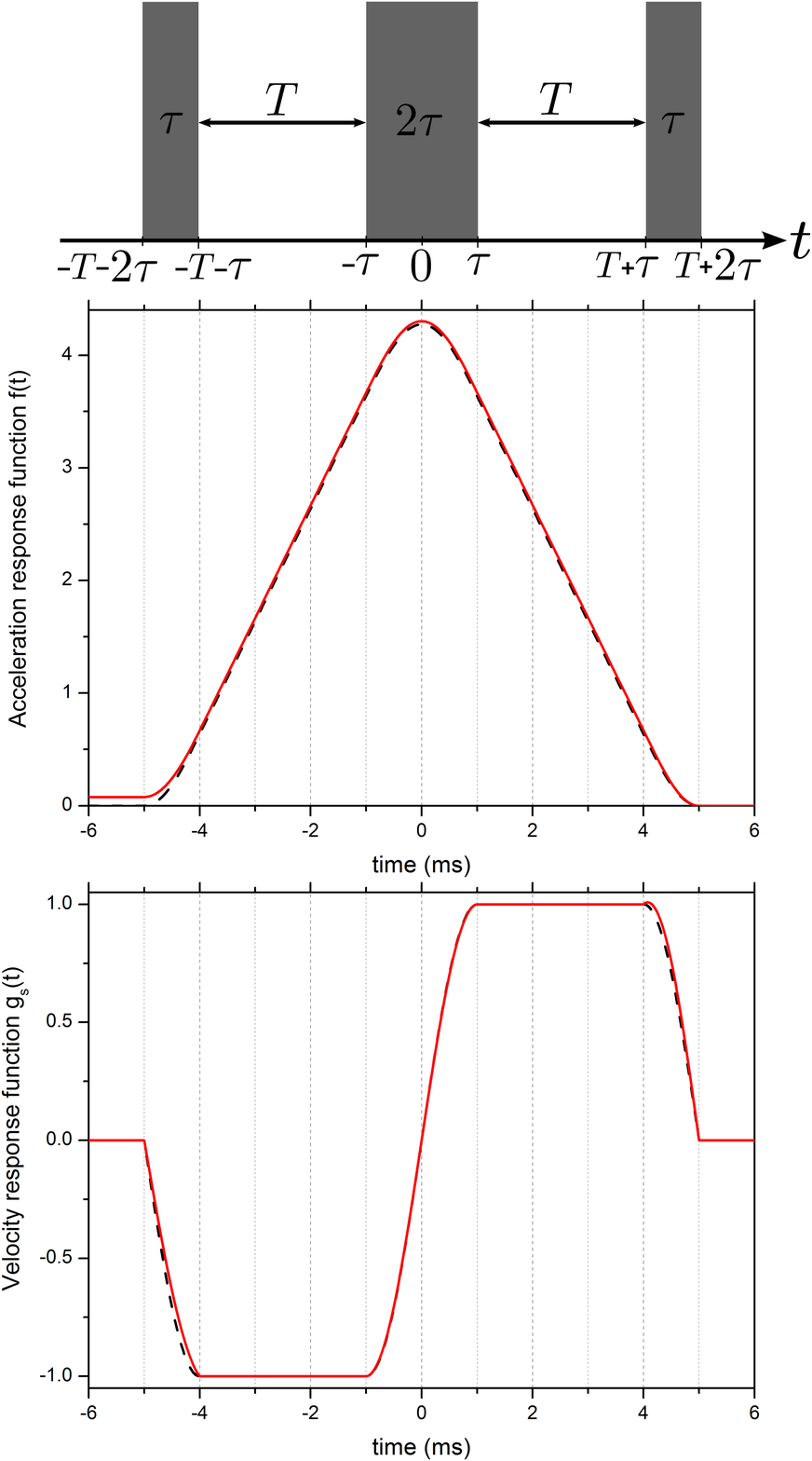}}
\caption
{
Mach-Zehnder type Atom interferometer time diagram and associated acceleration response functions $f(t)$ and velocity response functions $g_{s}(t)$. The black dashed line curves correspond to $\Omega_{1}=\Omega_{2}=\Omega_{3}=\frac{\pi}{2\tau}$, the red solid line curves correspond to $\Omega_{1}\tau=0.85\frac{\pi}{2}$, $\Omega_{2}2\tau=0.99\pi$ and $\Omega_{3}\tau=1.08\frac{\pi}{2}$ similar to values of the Rabi frequencies associated to $^{85}$Rb in our experiment. Here, $T=3$ ms and $\tau=1$ ms, the small value of $T$ and high value of $\tau$ are arbitrarily chosen for a good readability of the curve but have no connection with the reality of the experiment.
}
\label{time response function}
\end{figure}

The expression of the response function to acceleration is then given by:
\begin{equation}\label{response function}
f_{i}(t)=-\int^{t} g_{s}^{i}(t')\,dt'
\end{equation}
The Fourrier transform of this function gives the acceleration transfer function of the atom interferometer (see figure \ref{normalized transfer function}). The calculation and the exact expressions of both $g_{s}(t)$ and $f(t)$ functions are given in Appendix \ref{Appendix Response function}.

Two response functions, in acceleration and velocity, of a Mach-Zehnder type atom interferometer, corresponding to two different sets of Rabi frequencies ($\Omega_{1}=\Omega_{2}=\Omega_{3}$ and $\Omega_{1}\neq\Omega_{2}\neq\Omega_{3}$), are shown in FIG.\ref{time response function}. In the case where the Rabi frequencies perfectly match each other, the acceleration response function is rigorously even and is zero outside the light pulses sequence. This properties illustrate the fact that the interferometer is sensitive to acceleration during the interferometric phase only. In the case where the Rabi frequencies are different one to another, the acceleration response function still remains even on the interval $[-T-\tau , T+\tau]$ whereas the parity is broken between the first ($[-T-2\tau,-T-\tau]$) and the third ($[T+\tau,T+2\tau]$) pulses if $\Omega_{3}\neq\Omega_{1}$. It then follows that the function does not go to zero before the first Raman pulse, which is the signature of an interferometer sensitive to the atom initial velocity. When $\Omega_{3}>\Omega_{1}$, the acceleration response function is slightly above the response function corresponding to $\Omega_{1}=\Omega_{2}=\Omega_{3}$ and is thus positive for $t<-T-2\tau$ (\textit{cf.} FIG. \ref{time response function}).

Thanks to this formalism, the scale factor $S$ impacting the measurement of a constant acceleration can be derived from Eq.(\ref{sensitivity function and interferometric phase}). For deriving $S$, we consider that the velocity of the atoms evolves as $at$. $S$ is thus given by:

\begin{equation}\label{scale factor}
\begin{array}{lll}
S &=& k^{i}_{\scriptstyle\textrm{eff}}\int_{-\infty}^{-\infty} g_{s}^{i}(t)t\,dt \vspace{4pt}\\
 &\simeq& k^{i}_{\scriptstyle\textrm{eff}}(T+2\tau)(T+\frac{1}{\Omega_{1}}\tan(\frac{\Omega_{1}\tau}{2})+\frac{1}{\Omega_{3}}\tan(\frac{\Omega_{3}\tau}{2})) \vspace{4pt}\\
 &\simeq& k^{i}_{\scriptstyle\textrm{eff}} T^{2}
\end{array}
\end{equation}

\noindent This expression gives the measurement scale factor at first order in $\tau$ and $1/\Omega$. An experimental validation of this expression will be presented in section \ref{Difference in temporal response function} and the complete expression of the scale factor is given in appendix \ref{Appendix Response function} (\textit{cf.} Eq. (\ref{scale factor 2})). It is important to notice that at first order, the scale factor depends on Rabi pulsations of the first and third pulses only. It can be easily understood because of the interferometer symmetry. The atomic mirror efficiency will impact the contrast of the interferometer only whereas the atomic beam splitters efficiencies will impact the contrast and the phase of the interferometer.

We have now all the required tools to rigorously express the differential phase $\phi_{d}$. All the terms composing it are listed in Table \ref{differential phase table}. They can be gathered in five categories.

\begin{table*}
\caption
{
\label{differential phase table}
Phase terms composing the expression of the differential phase $\phi_{d}$ and gathered in five categories: WEP violation signal terms ($\Delta a$), mutual constant acceleration terms ($a$), vibration terms ($\tilde{a}$), velocity terms ($v$), and systematics. The second column presents phase shift terms according to the response and sensitivity functions formalism. The third column presents the dominant term of the expansion in the low frequency limit ($\omega \rightarrow 0$). The notations are mainly described in the text. For terms 1 to 10 the mismatch between $\Omega_{1}$ and $\Omega_{3}$ has been neglected and we consider that $\Omega_{1}=\Omega_{2}=\Omega_{3}$. This mismatch can not be neglected for velocity dependent terms (11 to 14) because they only appear for $\Omega_{1}\neq\Omega_{3}$, where $\frac{\delta\Omega^{i}}{(\Omega^{i})^{2}}$ refers to $\frac{1}{\Omega^{i}_{3}}\tan(\frac{\Omega^{i}_{3}\tau}{2})-\frac{1}{\Omega^{i}_{1}}\tan(\frac{\Omega^{i}_{1}\tau}{2})$ The fourth column gives the numerical size of the phase shift terms according to our experimental conditions ($T=47$ ms, $\tau=4~\mu$s, $a=g=9.81$ m.s$^{-2}$, $\Delta a=2\times10^{-14}$ m.s$^{-2}$, $\tilde{a}=3\times10^{-7}$ m.s$^{-2}$, $\Delta v = 6$ mm.s$^{-1}$, $v_{85}=0.6$ m.s$^{-1}$, $\delta k/k_{\scriptstyle\textrm{eff}}^{87}=5\times10^{-6}$, $|\int \delta f(t)\,dt|/|\int f_{87}(t)\,dt|=3\times10^{-6}$, $\delta \Omega/\Omega=0.1$ (between the first and the third Raman pulse and between both isotopes)).
}
\begin{ruledtabular}
\begin{tabular}{c c c c}

			\rule[0cm]{0pt}{0.2cm} Term & Phase Shift & Dominant Term & Size\\
			\rule[0cm]{0pt}{0.2cm}  &  &(DC term, $\omega \rightarrow 0$) & (rad)\\
			\hline
			
			\rule[0.2cm]{0pt}{0.2cm} $\Delta a$ Terms: &  & &\\
			
			\rule[0.2cm]{0pt}{0.2cm} \bfseries{1} & $k_{\scriptstyle\textrm{eff}}^{87} \Delta a \int f_{87}(t)\,dt$ & $k_{\scriptstyle\textrm{eff}}^{87} \Delta a (T+2\tau)(T+\frac{2}{\Omega^{87}}\tan(\frac{\Omega^{87}\tau}{2}))$ & $7\times10^{-12}$\\
			
			\rule[0.2cm]{0pt}{0.2cm} 2 & $\delta k \Delta a \int f_{87}(t)\,dt$ & $\delta k \Delta a (T+2\tau)(T+\frac{2}{\Omega^{87}}\tan(\frac{\Omega^{87}\tau}{2}))$ & $3.5\times10^{-15}$\\
			
			\rule[0.2cm]{0pt}{0.2cm} 3 & $k_{\scriptstyle\textrm{eff}}^{87} \Delta a \int \delta f(t)\,dt$ & $k_{\scriptstyle\textrm{eff}}^{87} \Delta a (T+2\tau)(\frac{2}{\Omega^{85}}\tan(\frac{\Omega^{85}\tau}{2})-\frac{2}{\Omega^{87}}\tan(\frac{\Omega^{87}\tau}{2}))$ & $2\times10^{-15}$\\

			\rule[0.2cm]{0pt}{0.2cm} 4 & $\delta k \Delta a \int \delta f(t)\,dt$ & $\delta k \Delta a (T+2\tau)(\frac{2}{\Omega^{85}}\tan(\frac{\Omega^{85}\tau}{2})-\frac{2}{\Omega^{87}}\tan(\frac{\Omega^{87}\tau}{2}))$ & $1\times10^{-20}$\\
			
			\rule[0.2cm]{0pt}{0.2cm} a Terms: &  & &\\
			
			\rule[0.2cm]{0pt}{0.2cm}  &  & &\\
			
			\rule[0.2cm]{0pt}{0.2cm} \bfseries{5} & $\delta k a \int f_{87}(t)\,dt$ & $\delta k a (T+2\tau)(T+\frac{2}{\Omega^{87}}\tan(\frac{\Omega^{87}\tau}{2}))$ & $1.73$\\
			
			\rule[0.2cm]{0pt}{0.2cm} 6 & $(k_{\scriptstyle\textrm{eff}}^{87} a -2\pi\alpha) \int \delta f(t)\,dt$ & $(k_{\scriptstyle\textrm{eff}}^{87} a -2\pi\alpha) (T+2\tau)(\frac{2}{\Omega^{85}}\tan(\frac{\Omega^{85}\tau}{2})-\frac{2}{\Omega^{87}}\tan(\frac{\Omega^{87}\tau}{2}))$ & $1\times10^{-7}$\\
			
			\rule[0.2cm]{0pt}{0.2cm} 7 & $\delta k a \int \delta f(t)\,dt$ & $\delta k a (T+2\tau)(\frac{2}{\Omega^{85}}\tan(\frac{\Omega^{85}\tau}{2})-\frac{2}{\Omega^{87}}\tan(\frac{\Omega^{87}\tau}{2}))$ & $4.8\times10^{-6}$\\
			
			\rule[0.2cm]{0pt}{0.2cm} $\tilde{a}$ Terms: &  & &\\
			
			\rule[0.2cm]{0pt}{0.2cm}  &  & &\\
			
			\rule[0.2cm]{0pt}{0.2cm} \bfseries{8} & $\delta k \int f_{87}(t)\tilde{a}(t)\,dt$ & $\delta k \tilde{a} (T+2\tau)(T+\frac{2}{\Omega^{87}}\tan(\frac{\Omega^{87}\tau}{2}))$ & $5.3\times10^{-8}$\\
			
			\rule[0.2cm]{0pt}{0.2cm} \bfseries{9} & $k_{\scriptstyle\textrm{eff}}^{87} \int \delta f(t)\tilde{a}(t)\,dt$ & $k_{\scriptstyle\textrm{eff}}^{87} \tilde{a} (T+2\tau)(\frac{2}{\Omega^{85}}\tan(\frac{\Omega^{85}\tau}{2})-\frac{2}{\Omega^{87}}\tan(\frac{\Omega^{87}\tau}{2}))$ & $3\times10^{-8}$\\
			
			\rule[0.2cm]{0pt}{0.2cm}10 & $\delta k \int \delta f(t)\tilde{a}(t)\,dt$ & $\delta k \tilde{a} (T+2\tau)(\frac{2}{\Omega^{85}}\tan(\frac{\Omega^{85}\tau}{2})-\frac{2}{\Omega^{87}}\tan(\frac{\Omega^{87}\tau}{2}))$ & $1.6\times10^{-13}$\\
			
			\rule[0.2cm]{0pt}{0.2cm} $v$ Terms  &  & &\\
			
			\rule[0.2cm]{0pt}{0.2cm}  &  & &\\
			
			\rule[0.2cm]{0pt}{0.2cm} \bfseries{11} & $k_{\scriptstyle\textrm{eff}}^{87} \Delta v \int g_{s}^{87}(t)\,dt$ & $k_{\scriptstyle\textrm{eff}}^{87} \Delta v \frac{\delta\Omega^{87}}{(\Omega^{87})^{2}}$ & $1.2\times10^{-2}$\\
			
			\rule[0.2cm]{0pt}{0.2cm} 12 & $ \left(k_{\scriptstyle\textrm{eff}}^{87} v_{85} +2\pi\alpha t_{r} \right) \int \delta g_{s}(t)\,dt$ & $\left(k_{\scriptstyle\textrm{eff}}^{87} v_{85} +2\pi\alpha t_{r} \right) (\frac{\delta\Omega^{85}}{(\Omega^{85})^{2}}-\frac{\delta\Omega^{87}}{(\Omega^{87})^{2}})$ & $1.2\times10^{-3}$\\
			
			\rule[0.2cm]{0pt}{0.2cm} 13 & $\delta k v_{85} \int g_{s}^{87}(t)\,dt$ & $\delta k v_{85} \frac{\delta\Omega^{87}}{(\Omega^{87})^{2}}$ & $6.1\times10^{-6}$\\
			
			\rule[0.2cm]{0pt}{0.2cm} 14 & $ \delta k v_{85} \int \delta g_{s}(t)\,dt$ & $\delta k v_{85} (\frac{\delta\Omega^{85}}{(\Omega^{85})^{2}}-\frac{\delta\Omega^{87}}{(\Omega^{87})^{2}})$ & $6.1\times10^{-7}$\\
			
			\rule[0.2cm]{0pt}{0.2cm} Systematics: &  & &\\
			
			\rule[0.2cm]{0pt}{0.2cm}  &  & &\\
			
			\rule[0.2cm]{0pt}{0.2cm} 15 & $\phi^{85}_{SE}-\phi^{87}_{SE}$ &  $\phi^{85}_{SE}-\phi^{87}_{SE}$ & -\\
\end{tabular}
\end{ruledtabular}
\end{table*}

Firstly, terms coming from a potential WEP violation signal. Term 1 is the WEP violation signal $\Delta a$ impacted by the measurement scale factor. Terms 2 and 3 are the first order corrections respectively coming from the differences in wave vectors, $\delta k = k_{\scriptstyle\textrm{eff}}^{85} - k_{\scriptstyle\textrm{eff}}^{87}$, and response functions, $\delta f(t) = f_{85}(t) - f_{87}(t)$, between both isotopes. Term 4 is the second order correction which combines both differences.

Secondly, terms depending on the mutual constant acceleration $a$ undergone by both species, which are not canceled because of potential differences between both scale factors. Term 5 corresponds to the differential phase shift induced by the constant acceleration because of the difference in wave-vectors and term 7 combines this effect with the difference in response functions. A difference in response function coupled to a middling correction of the Doppler shift by the Raman frequency chirp (\textit{i.e.} $2\pi\alpha\neq k_{\scriptstyle\textrm{eff}}^{87} a$) induces an additional phase shift (term 6).

Thirdly, terms depending on vibration noise $\tilde{a}(t)$ and which will cause some limitations of the common-mode vibration noise rejection. These limitations result from $\delta k$ (term 8), $\delta f(t)$ (term 9) or both (term 10).

Fourthly, when $\Omega_{1}\neq\Omega_{3}$ the interferometer becomes sensitive to a constant velocity of the atom along the Raman laser (reflected by the fact that $\int_{-\infty}^{+\infty} g_{s}^{i}(t)\,dt \neq 0$). This induces a differential phase shift if the vertical constant velocities associated to each isotope are not equal $\Delta v = v_{85}-v_{87}\neq0$ (term 11). It then exists some correcting terms coming from the difference in wave vectors (term 13 and 14) and in sensitivity functions $\delta g_{s}(t) = g_{s}^{85}(t) - g_{s}^{87}(t)$ (term 12 and 14). Term 12 depends also on the start time $t_{r}$ of the Raman chirp which is set as to correspond to the time when the atoms are released from the trap.

Fifthly, it obviously exists some systematics (term 15) which must be taken into account and corrected. In our experiment, the main systematics are the impact of additional laser lines, the two-photon light shift, the Coriolis effect and wave-front aberrations of the Raman laser \cite{Bonnin2013}.

\section{Differential phase extraction}
\label{Differential phase extraction}

The two signals from the dual-species atom interferometer are sinusoidal functions and thus parametrically describe an ellipse ($cf.$ Eq. (\ref{fringes expression})). In order to benefit from the coupling between both sensors few methods for deriving the differential phase from this ellipse have already been developed. The operating range of a simultaneous dual-species atom interferometer is much larger than for a ``standard" single-species atom gravimeter. Indeed, even when the acceleration fluctuations are greater than one fringe spacing, the ellipse remains visible and the differential acceleration can still be derived from it. This is the signature of the correlation between both interferometric signals from both isotopes which is made possible by the simultaneous aspect of the measurement.

A first solution, historically developed for gradiometers, consists in executing a least-square fitting method \cite{Foster2002}. This method permits a rapid extraction of the differential phase but is not bias-free in the presence of noises. A Bayesian analysis is a much more comprehensive approach for estimating the differential phase \cite{Stockton2007,Varoquaux2009} and will lead to an optimal estimator with negligible systematic error. This method can also be employed in cases where the scale factor associated to each species are different \cite{Chen2014}. Nevertheless, this method requires an accurate statistical model for the interferometer parameters and is computationally intensive to implement. For our data processing we have used an alternative method referred as ``Direct Phase Extraction'' \cite{Wu_thesis} that is described in this section.

In our case, the ellipse parameter corresponds to the common interferometric phase $\Delta\Phi_{87}$ shared by both isotopes which can be swept by scanning the Raman frequency ramp $\alpha$ or by introducing some vibration noise. The differential phase $\phi_{d}$ is related to the ellipticity, when $\phi_{d}=\pi/2$ the normalized ellipse is a circle, when $\phi_{d}=0$ the ellipse collapses to a line. More specifically, these curves are not closed Lissajous curves. Indeed, the frequencies of the two sine functions are slightly different because of the different scale factors related to each atomic species. Nevertheless this aspect is clearly negligible in our case, considering the slight difference in scale factors and the amplitude of vibrations. The parametric curves are thus approximated by ellipses in the paper. After each interferometric cycle, a couple of points ($x_{j}$,$y_{j}$) is obtained. The measurement is re-iterated $N$ times and N couples are finally obtained to describe the ellipse, such as:

\begin{equation}\label{ellipse parametric equation}
\left\{
	\begin{array}{ll}
			\vspace{0cm}
			x_{j} = -2\frac{P_{87,j}-P_{87,j}^{0}}{C_{87,j}} = \cos(\Delta\Phi_{87,j})
			\\
			y_{j} = -2\frac{P_{85,j}-P_{85,j}^{0}}{C_{85,j}} = \cos(\Delta\Phi_{87,j}+\phi_{d,j})
	\end{array}
\right.\
\end{equation}

From this equation, the common interferometric phase $\Delta\Phi_{87,j}$ can be ``directly" eliminated by using inverse functions (arcsin and arccos). By using some algebra, each couple ($x_{j}$,$y_{j}$) leads to two solutions of $\phi_{d}$ equal to

\begin{equation}\label{two solutions of phi d}
	\begin{array}{ll}
		\forall j \in [1,N]
		\\
		\phi_{d,j}=\arccos\left( x_{j}y_{j}\pm\sqrt{(1-x_{j}^{2})(1-y_{j}^{2})} \right), \in [0,\pi]
	\end{array}
\end{equation}

The final value of $\phi_{d}$ is extracted by using a maximum likelihood estimation over all $\phi_{d,j}$. Firstly, the phase estimation likelihood function, containing the $2N$ solutions, is fitted by a gaussian function to find its maximum in order to have a first estimation of $\phi_{d}$. According to this first estimation, for all $j$, only the closest value $\phi_{d,j}$ is then kept among the two solutions (\textit{cf.} Eq.(\ref{two solutions of phi d})). Indeed for each couple of point $j$, only one solution $\phi_{d,j}$ gives the correct value of $\phi_{d}$ on the interval $[0,\pi]$, the other being randomly distributed on $[0,\pi]$ when the ellipse is randomly parametrized. Finally, $\phi_{d}$ is given by the mean value of the second phase estimation likelihood function, now containing only N values $\phi_{d,j}$ associated to the N measurements ($x_{j}$,$y_{j}$). The one-sigma resolution is given by the standard deviation of the likelihood function over $\sqrt{N}$. In our experimental conditions the dominant noises are gaussian noises. The resulting likelihood function is thus symmetric and its mean value corresponds to the more likely value of the differential phase.

\begin{figure}
\centerline{\includegraphics[width=8cm]{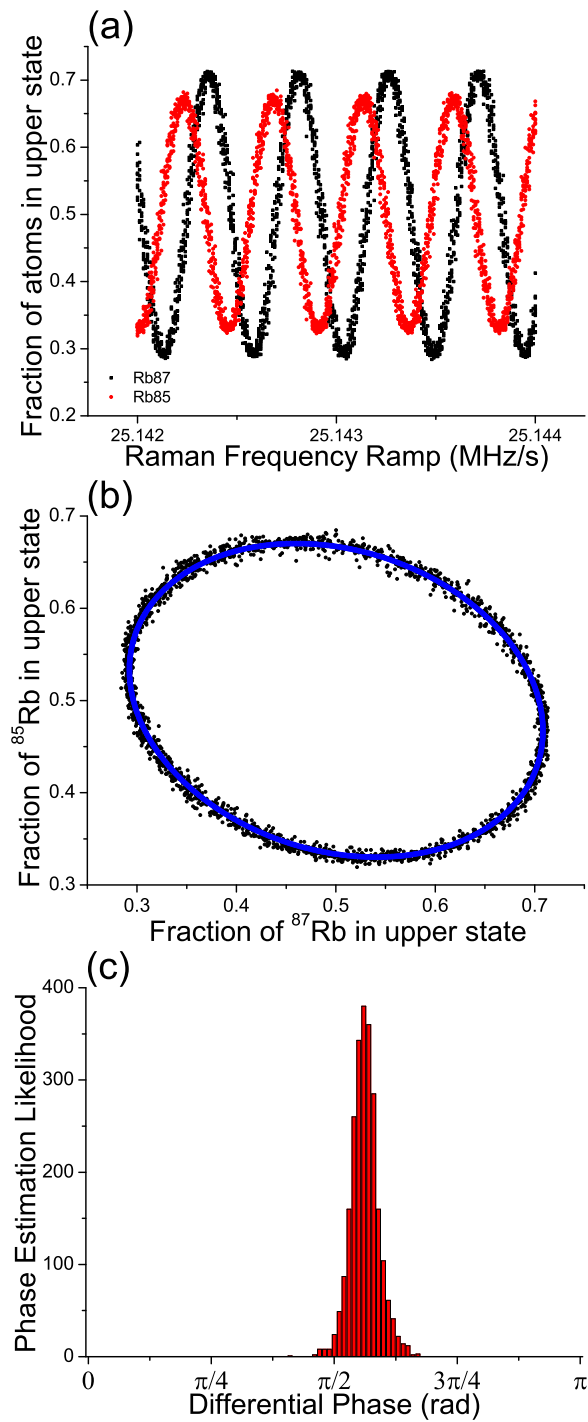}}
\caption
{
Simultaneous interferometric signals for $^{87}$Rb and $^{85}$Rb, with $T=47$ ms. From top to bottom, (a) interference fringes as a function of the microwave chirp $\alpha$, in black squares from $^{87}$Rb in red dots from $^{85}$Rb. (b) Signal from $^{85}$Rb is plotted vs signal from $^{87}$Rb drawing the ellipse fitted by the solid blue line. (c) $\phi_{d}$ is derived from the phase estimation likelihood function (red histogram). The fringes and the ellipse contain 2400 points, each point is acquired in 0.25 s corresponding to a global integration time of 600 s. The data were recorded with a single direction of the Raman wave-vector $+\vec{k}_{\scriptstyle\textrm{eff}}$.
}
\label{fringes ellipse and phase estimation}
\end{figure}           

Typical simultaneous dual-species fringes are reported in Fig. \ref{fringes ellipse and phase estimation}(a). Figure \ref{fringes ellipse and phase estimation}(b) shows the experimental ellipse obtained by plotting the interferometric signal from $^{85}$Rb versus the one from $^{87}$Rb. The phase estimation likelihood function ($cf.$ Fig. \ref{fringes ellipse and phase estimation}(c)) represents the differential phase distribution from this ellipse ($\phi_{d}\sim 1.7$ rad).

This differential phase extraction method is quick and easy to implement and chiefly allows for a bias-free differential phase extraction on a large range of $\phi_{d}$. We performed computing simulations in order to verify the bias-free aspect of this method. For different value of $\phi_{d}\in[0,\pi]$ and different amplitudes of noises, $\phi_{d}$ is estimated for a large number of noisy ellipses. This estimation method is bias-free on the range $[3\pi/16,13\pi/16]$ for an offset noise $\sigma_{P}/P$ of $1\%$, an amplitude noise $\sigma_{C}/C$ of $2\%$ and a non common phase noise of $\sigma_{\phi}=30$ mrad (all corresponding to noises three times larger than our experimental noises). This method requires an a priori knowledge of the fringe amplitudes ($C_{i}$) and population offsets ($P^{0}_{i}$). These parameters can be easily deduced by fitting the sinusoidal fringes, scanned either by the Raman frequency ramp $\alpha$ in a low vibration environment or by the vibrations themselves, where in that case, the atom sensor is correlated with a mechanical accelerometer. The probability density function \cite{Geiger2011} of the atom interferometer measurements can also give access to this parameters. A last method can be to simply derive the fringe amplitudes and offsets from the mean value and the standard deviation of the probability density function without any fitting method. No significant biases on the differential phase estimation appears for relative errors less than $5\%$ on the fringe contrasts and $1\%$ on the offsets. All the conducted numerical simulation tests prove thus the high reliability and robustness of this phase extraction method.

In a ``standard'' single-species atom gravimeter the phase resolution is proportional to the inverse of signal to noise ratio (SNR), \textit{i.e.} it scales as SNR$^{-1}$ \cite{Peters2001}. With the present Direct Phase Extraction method, this dependency is statistically estimated at SNR$^{-0.7}$. This behavior is empirically estimated by performing numerical simulations: for typical fixed values of offset noise, amplitude noise and non common phase noise, the SNR is modified by changing the fringes amplitudes. This phase extraction method is then less sensitive to acceleration fluctuations than the one used in a gravimeter \cite{Bidel2013}. This is here explained by the fact that the signal is integrated over all possible interferometric phases $[0,2\pi]$ and not only in the neighborhood of $\pi/2$ where the interferometer's response is the most sensitive.

\section{Experimental Apparatus}
\label{Experimental Apparatus}

For simultaneously measuring the acceleration undergone by $^{87}$Rb and $^{85}$Rb, both isotopes are first trapped and cooled thanks to the same laser beams and magnetic field gradients to form two spatially embedded magneto-optical traps (MOT). Both clouds are released from the trap and the accelerations are simultaneously measured by the same Mach-Zehnder type atom interferometer based on stimulated Raman transitions. 

The experimental setup is mainly derived from \cite{Bidel2013, Bonnin2013}. The cold atoms are trapped and cooled at the top of a high vacuum chamber made of glass and are then dropped over a distance of $\simeq$ 4 cm limiting $T$ to be smaller than 47 ms. The atoms are interrogated during their free fall by the Raman laser beam which is retro-reflected by a mirror representing the inertial reference for both isotopes. This mirror is attached to a mechanical accelerometer (nanometrics TITAN) to monitor its vibrations. The whole sensor head - containing the vacuum chamber, the magnetic shield consisting of four layers of mu-metal, the magnetic coils and the optics - is placed on a passive vibration isolation table (Minus-K). This table is itself mounted on an excitation table actuated by piezoelectric transducers (PZT). The isolation table resonance is tuned at the excitation frequency in order to filter higher harmonics to obtain a clean excitation at a well defined frequency.

The laser system for addressing both $^{87}$Rb and $^{85}$Rb is based on the frequency doubling of Telecom sources \cite{Carraz2009}. A distributed feedback (DFB) laser diode at 1560 nm is amplified  in a 5 W erbium-doped fiber amplifier (EDFA) and then frequency-doubled in a periodically poled lithium niobate (PPLN) crystal. The frequency of the laser is controlled thanks to a beat-note with a reference laser locked on a rubidium transition. The needed laser lines for the dual-species experiment are then synthesized thanks to a fibered electro-optic phase modulator at 1560 nm. Concerning the cooling and the detection stages, four laser lines are needed : cooling and repumping laser lines for both isotopes. In that cases, the carrier frequency is tuned on the $^{87}$Rb cooling transition. The three other laser lines are generated by injecting three microwave modulation frequencies (1.126 GHz, 2.915 GHz and 6.568 GHz) into the phase modulator as follows : 

\begin{equation}\label{MOT frequencies}
\begin{array}{lll}

f^{87}_{cooling}&=&f_{carrier}, \vspace{4pt}\\
f^{85}_{cooling}&=&f_{carrier} + 1.126~\mathrm{GHz}, \vspace{4pt}\\
f^{87}_{repumping}&=&f_{carrier} + 6.568~\mathrm{GHz}, \vspace{4pt}\\
f^{85}_{repumping}&=&f_{carrier} + 1.126~\mathrm{GHz} + 2.915~\mathrm{GHz}.

\end{array}
\end{equation}

\begin{figure}
\centerline{\includegraphics[width=8cm]{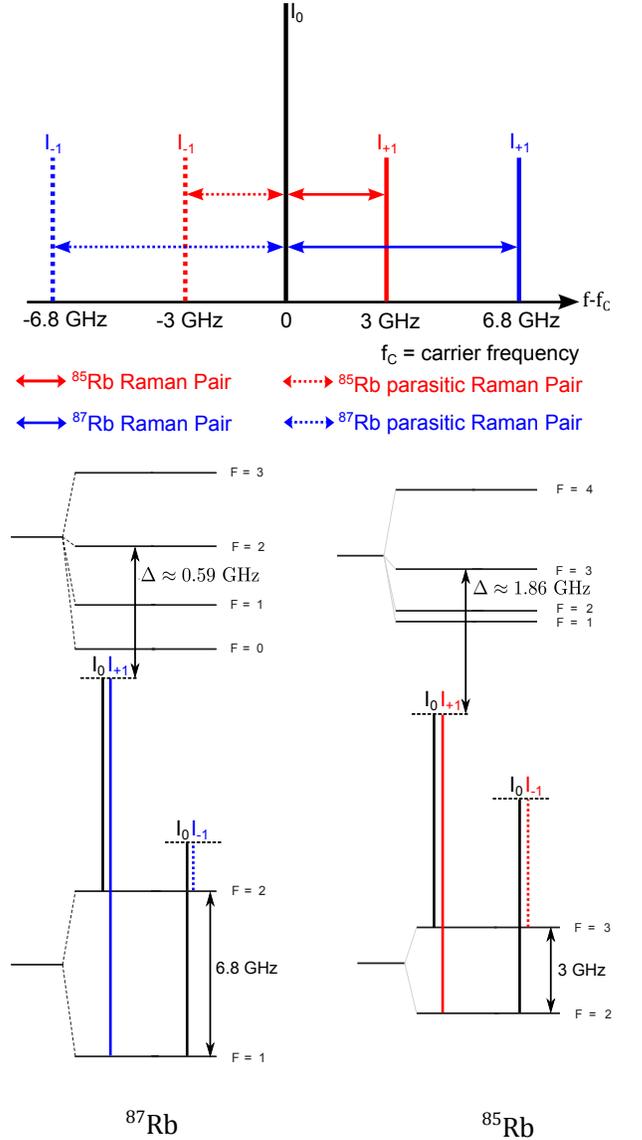}}
\caption
{
Schematic of the laser spectrum at 780 nm during the interferometric sequence with atomic level systems and Raman transitions for both isotopes. In solid lines the Raman pairs used for the atomic mirrors and beam splitters, in dashed lines the ``parasite" Raman pairs due to phase modulation generation, red color corresponding to $^{85}$Rb and blue to $^{87}$Rb.
}
\label{Raman laser spectrum}
\end{figure}

\noindent The power of each line is controlled by adjusting the modulation depth of each microwave frequency. During the cooling stage the experimental parameters are adjusted to obtain close trap features for both isotopes. Indeed both cooling line powers are set to be the same for both isotopes and the repumping power of $^{85}$Rb is a little stronger than the one of $^{87}$Rb, that compensates a higher depumping rate due to spontaneous emission because of its tighter hyperfine structure. Approximately one-third of the global laser power is lost in additional modulation lines far from any atomic resonances.  Phase modulation is also used for generating the laser lines during the interferometric sequence. Both Raman pairs are generated by directly injecting the Raman difference frequencies associated to each isotope (i.e., 6.834 GHz for $^{87}$Rb and 3.035 GHz for $^{85}$Rb), making the carrier frequency common to both Raman pairs (cf. FIG.\ref{Raman laser spectrum}). This way of laser frequencies generation leads to the creation of additional laser lines that can induce destructive interferences of the transition probability by driving ``parasites" Raman transitions \cite{Carraz2012} (cf. FIG.\ref{Raman laser spectrum}). The Raman pair corresponding to $^{87}$Rb is red-detuned by 0.59 GHz with respect to the excited hyperfine state $F'=2$ and therefore the one corresponding to $^{85}$Rb is red-detuned by 1.86 GHz with respect to $F'=3$. The power of these two pairs is adjusted to obtain Rabi frequencies of the two photons Raman transitions as identical as possible.

With this setup, few 10$^{8}$ atoms are loaded from a background vapor into MOTs in 250 ms. The atoms are then further cooled down in an optical molasses phase of 28 ms leading to a temperature of 1.6 $\mu$K for $^{87}$Rb and 2.6 $\mu$K for $^{85}$Rb. Additional trap loss collisions due to inter-species atomic collisions \cite{Suptitz1994} do not exceed 10-15\% in our case. These results confirm that the additional laser lines do not have any significant impact on the cooling efficiency. Atoms are furthermore selected in the Zeeman sub-level $m_{F}=0$ of the hyperfine ground state (i.e. $F=1$ for $^{87}$Rb and $F=2$ for $^{85}$Rb) to remain insensitive to parasites magnetic fields. The selection is made thanks to a microwave $\pi$-pulse which drives the transition between the two hyperfine ground states for atoms in the sub-level $m_{F}=0$ only. A slightly blue-detuned optical pulse removes any residual atomic population. 

During the free fall, the interferometric sequence occurs in a vertical uniform magnetic field of 28 mG. The sequence consists in three Raman laser pulses of durations $\tau-2\tau-\tau$ ,with $\tau=4$ $\mu$s, equally spaced in time by $T$ = 47 ms and realizing the $\pi/2$-$\pi$-$\pi/2$ sequence of the Mach-Zehnder type interferometer (cf FIG.\ref{time response function}). The Raman laser pulses couple at the same time the states $|F\!=\!1,m_{F}\!=\!0\!>$ to $|F\!=\!2,m_{F}\!=\!0\!>$ for $^{87}$Rb and $|F\!=\!2,m_{F}\!=\!0\!>$ to $|F\!=\!3,m_{F}\!=\!0\!>$ for $^{85}$Rb. Exactly the same microwave chirp $|\alpha| \simeq 25.143$ MHz.s$^{-1}$ is applied to both Raman difference frequencies in order to compensate the time-dependent Doppler shift induced by gravity (\textit{cf.} eq. (\ref{interferometer phase})). This chirp is synthesized with the same DDS (Digital Direct Synthesizer) for both isotopes.

Finally, the atomic population repartition between the two coupled states is measured for each species by fluorescence detection. The atomic cloud is illuminated by a vertical resonant detection beam and the fluorescence is collected thanks to a collimation lenses system and a photodiode on the perpendicular direction. The atomic cloud is illuminated by two successive sequences of three light pulses of durations 1.5-0.05-1.5 ms. The first sequence induces the fluorescence signal from $^{87}$Rb atoms: the first pulse detects atoms in $F=2$, the second one fully transfers atoms from $F=1$ to $F=2$, the third pulse is identical to the first one and detects atoms initially in $F=1$. During the first and third pulse the laser frequency is slightly blue-detuned (+0.88 $\Gamma$, where $\Gamma$ is the natural line-width of the transition) and no modulation is applied to detect and push away the atoms. During the repumping pulse (second one) the 6.568 GHz modulation is injected into the phase modulator to generate the repumping line with a power adjusted in order to suppress as much as possible the carrier frequency. The second sequence of three pulses induces the fluorescence signal from $^{85}$Rb and is conceptually identical to the first one: the first pulse detects atoms in $F=3$, the second one fully transfers atoms from $F=2$ to $F=3$, the third pulse is identical to the first one and detect atoms initially in $F=2$. In practice, during the first and third pulse the 1.126 GHz modulation is injected into the phase modulator to generate the cycling transition with a blue detuning of 0.7 $\Gamma$ and with a power allowing to cancel the carrier frequency and thus minimize a cross signal from $^{87}$Rb atoms. During the repumping pulse (second one) the 1.126 and 2.915 GHz modulations are injected into phase modulator to generate the repumping line. In that case, both cycling transitions can not be canceled. The fluorescence of the background vapor is finally recorded and removed for both isotopes. With this detection scheme, a detection noise on the transition probability of $\sigma_{P}=0.0027$ for $^{87}$Rb, and $\sigma_{P}=0.0035$ for $^{85}$Rb, is obtained.

The whole sequence is performed at a repetition rate of 4 Hz.

\section{Previous results \& improvements}
\label{Previous results and improvements}

Before giving the last results of the dual-species atom interferometer, we recapitulate the previous results presented in \cite{Bonnin2013} and the improvements that have followed.

The resolution on the differential phase was about $3\times10^{-8}g$ after an integration time of 15 min corresponding to a sensitivity of $1\times10^{-6}g/\sqrt{Hz}$. These performances were limited by the detection noise of the experiment. The fringe amplitude was about 22 \% for $^{87}$Rb and 7 \% for $^{85}$Rb mostly because of spurious impact of additional laser lines generated by modulation and of our detection scheme. Thanks to the simultaneous aspect of the experiment, we highlighted a common-mode vibration noise rejection between both isotopes higher than 55 dB (rejection factor of 550) limited by the detection noise. Moreover, we performed a quantum based test of the WEP at a level of few $10^{-7}$ limited by uncertainties over systematics. This experiment was the first demonstration of a simultaneous dual-species atom interferometer.

The first step to improve the experiment was to reduce the impact of additional laser lines generated by phase modulation. The spurious impact on the fringe visibility is briefly presented in appendix \ref{Appendix additional laser lines}, and a more comprehensive theoretical paper \cite{Carraz2012} studies the additional lines impacts more generally. The main point to emphasize is the spatial dependency of the probability amplitude between the two states coupled by the light. The atomic beam splitters and mirrors efficiency directly depends on this probability amplitude (given by the Rabi frequency $\Omega$). As atoms are in free fall, the two-photon Rabi frequency depends on the time when the light pulse occurs and thus the transfer efficiency depends on T. T is set at 47 ms to achieve the best sensitivity allowed by our experimental set-up and the retro-reflecting Raman mirror position has been adjusted to obtain Rabi frequencies for each pulse, and for each isotope, as close as possible. For instance in the current experiment the different Rabi frequencies are $\Omega^{87}_{1}=\pi/(7.7 \mu s)$, $\Omega^{87}_{2}=\pi/(8 \mu s)$, $\Omega^{87}_{3}=\pi/(7.9 \mu s)$, $\Omega^{85}_{1}=\pi/(9.4 \mu s)$, $\Omega^{85}_{2}=\pi/(8.1 \mu s)$, $\Omega^{85}_{3}=\pi/(7.4 \mu s)$, that is why the pulse duration $\tau=$ 4 $\mu$s has been chosen for realizing a Mach-Zehnder type atom interferometer $\pi/2-\pi-\pi/2$.

The non-common phase noise between both isotopes was reduced by optimizing the micro-wave source used for generating the Raman pair of $^{85}$Rb. Mainly, a lower noise frequency reference at 10 MHz was implemented in the optimized micro-wave source.

The detection has also been improved in order to increase the SNR. The interferometric signal (cf. Eq.\ref{fringes expression}) is given by the proportion of atoms $P$ in the upper hyperfine ground state at the output of the interferometer :
\begin{equation}\label{fringes and detection} 
	P = \frac{N_{2}}{N_{1}+N_{2}}
\end{equation}
where $N_{2}$ is the number of atoms in the upper state and $N_{1}$ in the lower one. The detection aims to accurately estimate the ratio $P$ by counting the atoms in each state. We name here $N_{1}^{det}$ and $N_{2}^{det}$ the detected signal associated to each state. Our detection scheme shows non-linearities when $N_{1}$ is measured (\textit{i.e.} $N_{1}^{det}$ is not proportional to $N_{1}$). The interferometric fringes are not perfectly sinusoidal any more and it results a bias on the determination of the differential phase from the ellipse. These non-linearities have been taken into account as in the following. 

The detection sequence has been explained in section \ref{Experimental Apparatus}. During the first detection pulse, when is $N_{2}$ measured, a part of these atoms, $\epsilon_{\mathrm{sp}}N_{2}$, are transferred to the lower state because of the depumping rate due to spontaneous emission ($0\leq\epsilon_{\mathrm{sp}}\leq1$). Moreover, some atoms, $\epsilon_{2}N_{2}$, are not perfectly pushed away from the detection area because of the finite duration of the pulse ($0\leq\epsilon_{2}\leq1$). These effects counterbalance each other (when $\epsilon_{2}$ increases, $\epsilon_{\mathrm{sp}}$ decreases and inversely) as a function of the detection laser power. When the laser intensity is low compare to the saturation intensity of the detection transition, a very small part of atoms are lost because of spontaneous emission whereas a huge part of atoms are not pushed away from the detection area, and inversely at high laser intensity.

During the second detection pulse, when atoms in the lower state are re-pumped in the upper state, it is impossible to completely cancel the blue detuned detection light. This residual pushes away a part of the atoms, $(1-\epsilon_{1})N_{1}$, from the detection area ($0\leq\epsilon_{1}\leq1$).

Consequently, during the third detection pulse, $N_{1}^{det}\propto\epsilon_{1}N_{1}+\epsilon_{1}(\epsilon_{\mathrm{sp}}+\epsilon_{2})N_{2}$ atoms will be detected instead of $N_{1}$. This effects must be evaluated corrected. The atomic population repartition can then be reconstruct by
\begin{equation}\label{fringes and detection and non linearities} 
	P = \frac{N_{2}^{det}}{\frac{N_{1}^{det}}{\epsilon_{1}}+(1-\epsilon_{\mathrm{sp}}-\epsilon_{2})N_{2}^{det}}
\end{equation}
Interferometric fringes shown in Fig. \ref{fringes ellipse and phase estimation}(a) are corrected from these non-linearities. We measured for $^{87}$Rb, respectively for $^{85}$Rb, $\epsilon_{1}=0.75$ and $\epsilon_{\mathrm{sp}}+\epsilon_{2}=0.12$, respectively $\epsilon_{1}=0.8$ and $\epsilon_{\mathrm{sp}}+\epsilon_{2}=0.3$. Micro-wave power fluctuations give rise to temporal fluctuations of these non-linearity coefficients. These temporal variations are taken into account by fitting the interferometric fringes and extracting the parameters $\epsilon_{1}$ and $\epsilon_{\mathrm{sp}}+\epsilon_{2}$ at different moments of the experiment.

The SNR is now approximately five times better than before \cite{Bonnin2013}, the fringe amplitudes being currently about 40\% for $^{87}$Rb and 35\% for $^{85}$Rb. This disparity is explained by the impact of the additional laser lines which still remains a bit stronger for $^{85}$Rb and by the visibility loss induced by the higher temperature of $^{85}$Rb and the velocity selection of stimulated Raman transitions \cite{Kasevich1991}.

\section{Resolution, sensitivity and long term stability}
\label{Resolution, sensitivity and long term stability}

The interferometric signals shown in Fig. \ref{fringes ellipse and phase estimation}, associated to a single direction of $\vec{k}_{\scriptstyle\textrm{eff}}$, result from a measurement acquisition of 600 s at a repetition rate of 4 Hz. By extracting the differential phase from the ellipse as explained in section \ref{Differential phase extraction}, $\phi_{d}$ is derived with a resolution of 1.5 mrad corresponding to a resolution of $5.10^{-9}g$ and a sensitivity of 1.23$\times$10$^{-7}g$ at 1 s, with our differential accelerometer by assuming that we are limited by a white noise.

\begin{figure}
\centerline{\includegraphics[width=10cm]{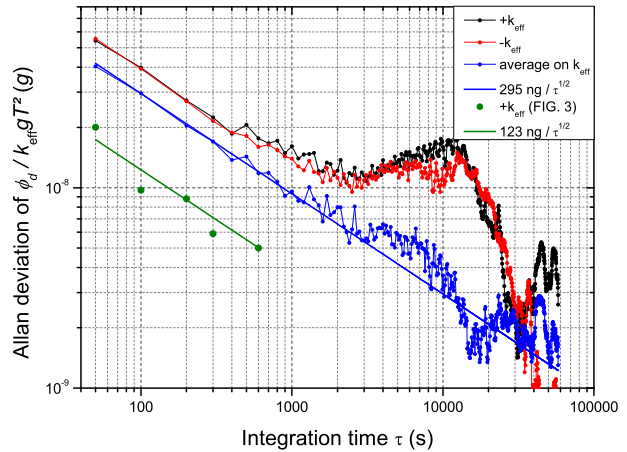}}
\caption
{
Allan deviations on the differential phase $\phi_{d}$ expressed in $g$ unit $\phi_{d}/(k_{\scriptstyle\textrm{eff}}^{87}gT^{2})$ with $g\simeq9.81$ m.s$^{-2}$. In black and red dots the Allan deviation corresponding to the two reversed directions of $\vec{k}_{\scriptstyle\textrm{eff}}$. The Allan deviation of the averaging over these two directions (blue dots) and its asymptotic behavior (blue line) is fitted by 295 n$g$/$\sqrt{\tau}$. The best sensitivity achieved with our differential accelerometer (green dots, corresponding to data points from FIG. \ref{fringes ellipse and phase estimation}) and its asymptotic behavior (green line) is fitted by 123 n$g$/$\sqrt{\tau}$. The first data point at $\tau=50$ s is given by the time required to obtain enough points to extract the phase from the ellipse (here 100 points).  
}
\label{Allan deviation}
\end{figure}

In order to study the long term stability of the instrument, we alternatively record at 4 Hz the signal for the direction $+\vec{k}_{\scriptstyle\textrm{eff}}$ and then for the opposite one $-\vec{k}_{\scriptstyle\textrm{eff}}$. A value of the differential phase $\phi_{d}$ is thus obtain every 0.5 s. The direction of $\vec{k}_{\scriptstyle\textrm{eff}}$ is changed drop by drop to eliminate some systematic effects (mainly the one photon light shift and the first order Zeeman shift) whose sign does not change with $\vec{k}_{\scriptstyle\textrm{eff}}$. We split the atom interferometer data points into groups of 100 consecutive points as a minimum of 100 points is required to derive a confident value of $\phi_{d}$ with our phase extraction method. The interferometric phase $\Delta\Phi_{87}$ is randomly distributed over several fringes thanks to vibrations as the vibration isolation system was turned off during the data acquisition. For each group composed of 100 points, a value and a standard error are derived for $\phi_{d}$. For each group of points, the estimation of detection non-linearities parameters as well as the estimations of the fringe amplitudes and population offsets are realized by fitting the interferometric fringes. This fit is achievable thanks to the correlation between the atomic and the mechanical accelerometer that allows to reconstruct the interferometric fringes.

Figure \ref{Allan deviation} shows the Allan deviation on the differential phase $\phi_{d}$, expressed in $g$ unit ($\phi_{d}/(k_{\scriptstyle\textrm{eff}}^{87}gT^{2})$), where $g$ refers to the gravity acceleration ($g\simeq 9.81$ m.s$^{-2}$), associated with three possible set of points: $+\vec{k}_{\scriptstyle\textrm{eff}}$ points only, $-\vec{k}_{\scriptstyle\textrm{eff}}$ points only or averaging over the two opposite directions. This averaging allows to reject the long term drifts induced by the one-photon light shift fluctuations, indeed the Allan deviation does not contain the bump around 3 hours induced by these fluctuations.

These Allan deviation behaviors prove also that the short term sensitivity is limited by white noises, namely the detection noise and the non-common phase noise between both isotopes. In our experiment, the detection noise is induced by the laser frequency noise during the detection. This noise is estimated to limit the sensitivity at a level of 1.4$\times$10$^{-7}$$g$ at 1 s. The other noise source is the non-common phase noise induced by the phase noise of the two different micro-wave sources used for generating both Raman pairs by phase modulation. By experimentally measuring the power spectral density of phase noise associated to each source, we estimated that the impact of this noise is lower than 0.65$\times$10$^{-7}$$g$ at 1 s. We finally estimated a global white noise level limiting the sensitivity to 1.55$\times$10$^{-7}$$g$ at 1 s. This value is in good agreement with the sensitivities measured in FIG. \ref{Allan deviation}. 

The best experimental sensitivity (1.23$\times$10$^{-7}$$g$ at 1 s) was achieved with data points from FIG. \ref{fringes ellipse and phase estimation} for a single direction of $\vec{k}_{\scriptstyle\textrm{eff}}$. The sensitivity after the averaging over the opposite directions of $\vec{k}_{\scriptstyle\textrm{eff}}$ (2.95$\times$10$^{-7}$$g$ at 1 s, $cf.$ blue points in FIG.\ref{Allan deviation}) comes from an different set of data. During this measurement session, the detection was not fully optimized degrading the sensitivity.

The results shown in FIG. \ref{Allan deviation} demonstrate that our differential accelerometer has a sufficient resolution, by integrating the signal during only 5 hours, for testing the WEP at a level of 2$\times$10$^{-9}$.

\section{Vibration Noise Rejection}
\label{Rejection of vibration noise}

\subsection{Experimental Results}
\label{Experimental Results}

In our experiment, the differential acceleration measurement is performed with a simultaneous interrogation of both atomic species falling in a common reference frame. The simultaneous aspect allows us to benefit from an efficient common-mode noise rejection which is not possible when atoms are alternatively handled.

The sensitivity of most state-of-the-art gravimeter is limited by the vibration noise whereas the differential atom sensors intrinsically allows to efficiently reject it. The rejection of this noise, in an atom sensor, has already been studied in a gravity gradiometer \cite{McGuirk2002} handling two cold atomic clouds of the same species separated in space. Our experiment \cite{Bonnin2013} has demonstrated for the first time the rejection of vibration noise with two embedded clouds of $^{87}$Rb and $^{85}$Rb simultaneously submitted to the same light pulse interferometric sequence. In our case, this kind of rejection can be extended to other environmental perturbations, such as gravity gradients or rotations, which depend on the overlap of the two atomic ensembles.

The vibration noise rejection is a critical point for a sensor which might be used on a moving platform or in a highly noisy environment. In the more specific context of a WEP test, it is necessary that the vibration noise does not limit the sensitivity of the instrument, \textit{i.e.} terms 8, 9, 10 presented in Table \ref{differential phase table} must remain lower than the targeted one shot resolution. So the vibration noise rejection ratio must be evaluated in order to determine the vibration level limit at which the instrument still benefit from an adequate sensitivity.

\begin{figure}
\includegraphics[width=7cm]{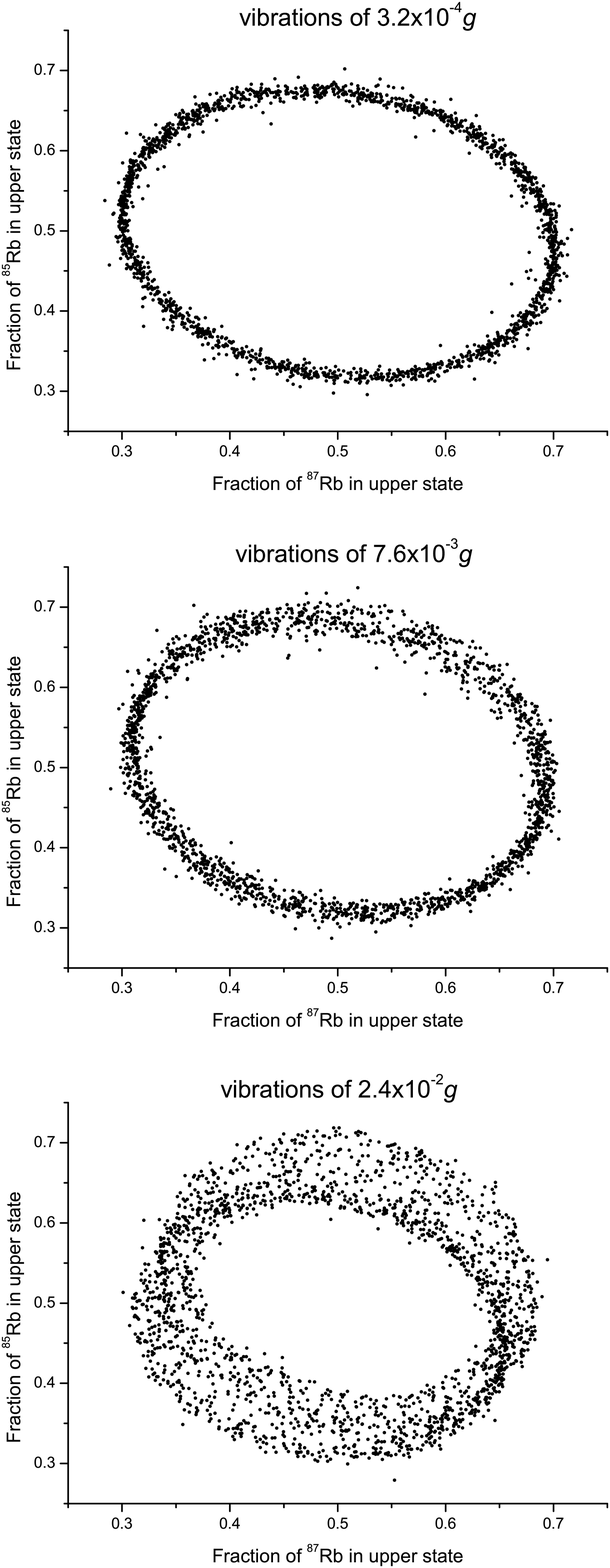}
\caption
{
Three ellipses of 2000 points plotted for three different vibration amplitudes: $3.2\times10^{-4}g$, $7.6\times10^{-3}g$ and $2.4\times10^{-2}g$. The ellipses are progressively blurred because of additional inertial effects induced by the excitation (cf. section \ref{Drop of contrast because of inertial effects}).
}
\label{3 ellipses}
\end{figure}

We characterized the rejection of vibrations by shaking the whole sensor, including the sensor head and the passive isolation platform, thanks to an excitation platform actuated by PZT. A first test was made by tuning the minus-K vertical resonance at a frequency of 2.08 Hz, identical to the excitation frequency. This setup allows to obtain a pure sinusoidal excitation at a given frequency. The 2.08 Hz excitation frequency is chosen to stand within the interferometer bandwidth ($0 \rightarrow 1/2T \simeq 0 \rightarrow 10$ Hz), to ensure that the demonstrated rejection comes from the differential measurement and not from the natural interferometer filter. Moreover this frequency is chosen to avoid aliasing effects with the experimental repetition rate of 4 Hz. Then, for different vibration amplitudes $A_{\scriptstyle\textrm{vib}}$ (from few 10$^{-5}g$ to few 10$^{-2}g$) the differential acceleration is derived by fitting ellipses containing $N=2000$ points. Figure \ref{3 ellipses} shows three of these ellipses for vibration amplitudes of $3.2\times10^{-4}g$, $7.6\times10^{-3}g$ and $2.4\times10^{-2}g$. Between each ellipse acquisition, the fringes amplitudes and the population offsets are estimated by fitting the interferometric fringes obtained thanks to the correlation between the atom and the mechanical accelerometer.

\begin{figure}
\centerline{\includegraphics[width=10cm]{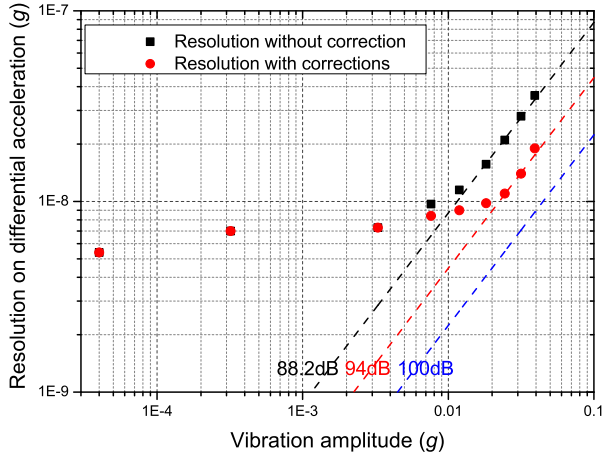}}
\caption
{
Resolution on the differential acceleration versus the amplitude of vibrations. The resolution is estimated from the ellipses, containing $N=2000$ points, for different amplitudes of vibrations at 2.08 Hz. The correlation with a mechanical accelerometer allows to measure the vibrations, and to estimate the fringe amplitudes and the population offsets. These two parameters depends on inertial effects ($cf.$ \ref{Drop of contrast because of inertial effects}), these dependency is (red dots) or is not (black squares) taken into account. In dashed lines, the resolution is plotted according different values of the rejection ratio $r=88.2$ dB (black), $r=94$ dB (red, maximum achieved with our experiment), $r=100$ dB (blue, theoretical limit with our experiment).
}
\label{vibration rejection}
\end{figure}

The results concerning the characterization of the vibration noise rejection are shown in Fig. \ref{vibration rejection}. When the amplitude of vibration is lower than typically few m$g$, the atom interferometer sensitivity is still limited by the detection noise and remains nearly constant over this range. Conversely, when the amplitude is higher than few m$g$, the vibrations limit the sensitivity and therefore the resolution becomes proportional to their amplitude. Under these experimental conditions the vibration rejection ratio $r$, defined as
\begin{equation}\label{rejection ratio} 
r = 20.\log\left(\frac{A_{\scriptstyle\textrm{vib}}}{\sqrt{N}\times \sigma_{\Delta g}}\right)
\end{equation}
can be experimentally evaluated at 88 dB, where $\sigma_{\Delta g}$ is the resolution on the differential acceleration. In addition to the resolution losses, the vibrations induce a drop of contrast through the Doppler shift because of additional vertical acceleration. Some contrast losses and offset population modifications come from rotations (through Coriolis effect \cite{Lan2012}, angular and centrifugal acceleration). These rotations are due in our case to a non purely vertical excitation with our platform. These effects are measured, during the measurement session thanks to the mechanical accelerometer and additional rotation sensors, and are post-corrected during the differential phase extraction from the ellipse (\textit{cf.} FIG.\ref{post corrected ellipse}). Taking these inertial effects into account, the rejection of vibration then reaches a level of 94 dB (a factor 50 000) between both isotopes. This vibration rejection level, obtained with two different species, is extremely encouraging, and chiefly demonstrates the robustness and reliability of a simultaneous differential acceleration measurement with $^{87}$Rb and $^{85}$Rb for testing the WEP in various environmental conditions.


In the next sections we will study what are the limits that can explain this level of vibration rejection.

\subsection{Limit - Wave-vector mismatch}

The first limit comes from the wave-vector mismatch between both species. This difference makes both scale factors slightly different and any spurious acceleration signals are not perfectly canceled by the differential measurement. This is illustrated by terms 7, 8 and 10 in Table \ref{differential phase table}. In our experiment $\delta k/k_{\scriptstyle\textrm{eff}}=5\times10^{-6}$, limiting the rejection ratio at the same level $i.e.$ 106 dB. This mismatch is dictated by the way in which the laser lines are generated. Indeed, by looking back at the Raman laser spectrum (Fig. \ref{Raman laser spectrum}), the fact that the carrier frequency is common to both Raman pairs makes impossible an equalization of wave-vectors.

\begin{figure}
\centerline{\includegraphics[width=10cm]{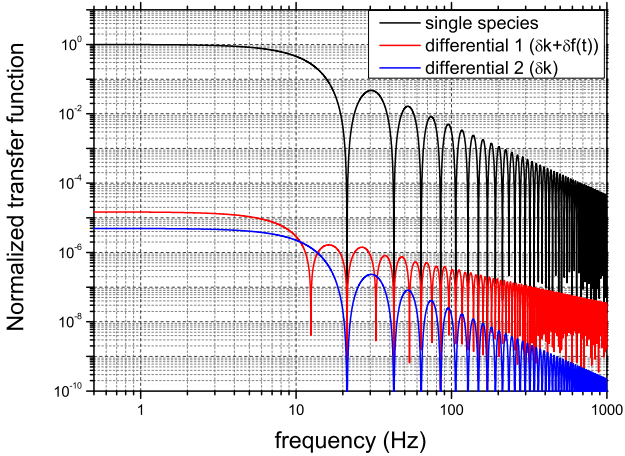}}
\caption
{
Normalized acceleration transfer function of a single species atom interferometer (black line, single species), a dual-species atom interferometer with wave-vector and Rabi frequencies mismatches (differential 1, red line), a dual-species atom interferometer with a wave-vector mismatch only (differential 2, blue line). The wave vector mismatch is $\delta k/k_{\scriptstyle\textrm{eff}}=5\times10^{-6}$, The Rabi frequencies mismatch between $^{87}$Rb and $^{85}$Rb is 0.1 (10\%).
}
\label{normalized transfer function}
\end{figure}

Figure \ref{normalized transfer function} displays the acceleration transfer function of a differential dual-species atom interferometer with a wave-vector mismatch (differential 2 curve) compare to the one of a single species interferometer (single species curve).

\subsection{Limit - Difference in temporal response functions (Rabi frequencies mismatch)}
\label{Difference in temporal response function}

Another source of difference between both scale factors is due to the difference of temporal response functions, $\delta f(t) = f_{85}(t) - f_{87}(t)$, to accelerations. In our experiment, the atomic beam splitters and mirror are realized by the same laser beam for both isotopes. That is why the pulse duration, $\tau$, and the free evolution time, $T$, will be inherently perfectly matched. Therefore, the response function difference comes only from the light-atom interaction making the scale factor dependent on the Rabi frequency associated to each Raman transition. That is why a Rabi frequency mismatch will limit the vibration rejection, it is illustrated by term 9 and 10 in Table \ref{differential phase table}.

\begin{figure}
\centerline{\includegraphics[width=10cm]{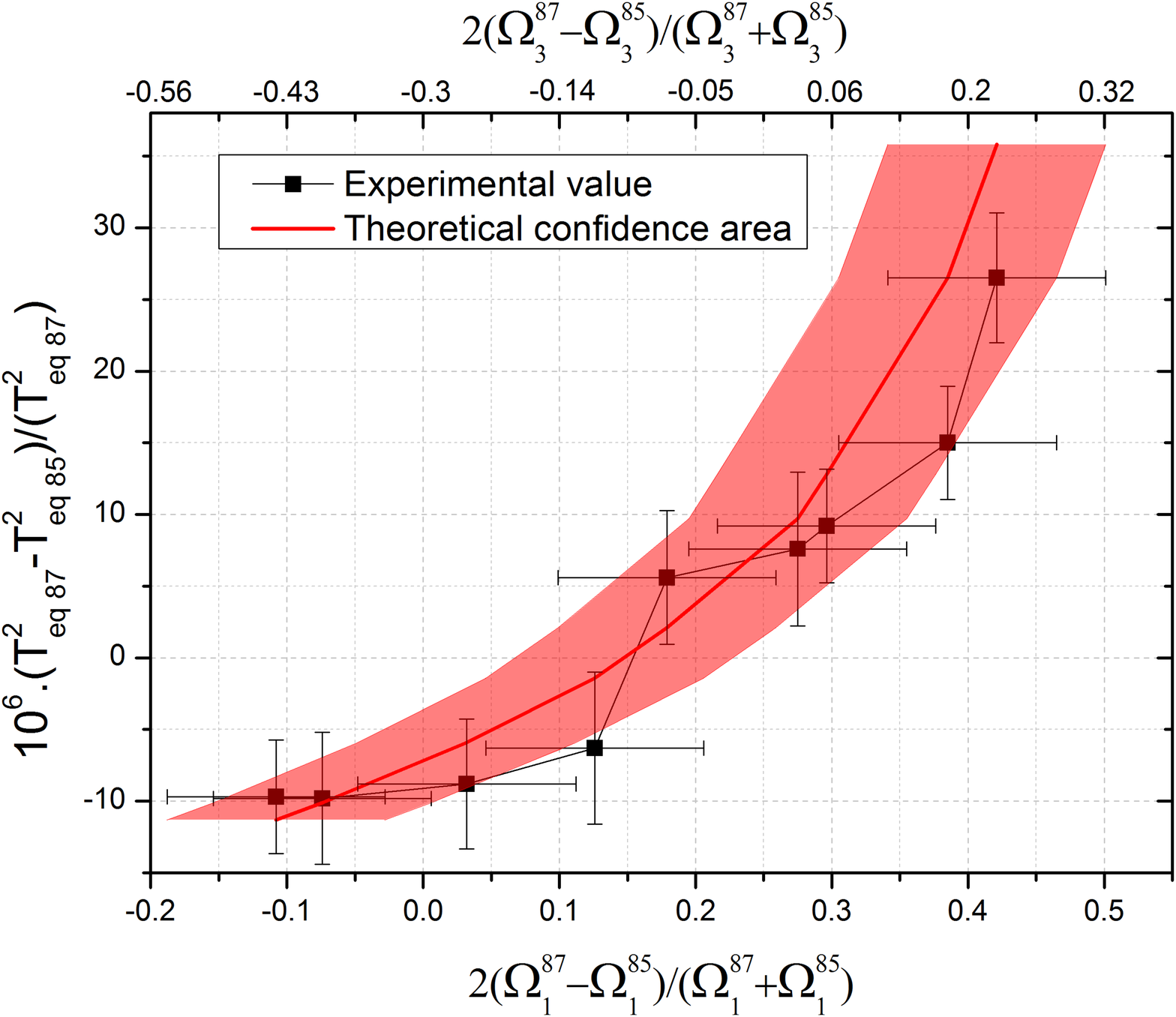}}
\caption
{
Difference in response functions $\int f_{87}(t)\,dt-\int f_{85}(t)\,dt=\int \delta f_{87}(t)\,dt=T^{2}_{eq,87}-T^{2}_{eq,85}$, in the low frequency limit, between $^{87}$Rb and $^{85}$Rb as a function of the Rabi frequency mismatch during the first and third pulses. The experimental measurement (black square) is compared to the theoretical value (red line) as expressed in Eq. (\ref{scale factor}). The horizontal uncertainty corresponds to the uncertainty on the determination of the Rabi frequency. The vertical uncertainty corresponds to the 1-$\sigma$ uncertainty coming from the determination of the scale factor by fitting the interferometric fringes.
}
\label{difference in response functions}
\end{figure}

In order to emphasize this limitation, we experimentally compared the difference in response function, in the low frequency limit, between both isotopes. For this measurement, a very large number of fringes is recorded ($\simeq$ 500) for both isotopes by sweeping the frequency chirp $\alpha$. The scale factors are extracted thanks to a sinus fitting, and the difference $\int f_{87}(t)\,dt-\int f_{85}(t)\,dt=\int \delta f(t)\,dt=T^{2}_{eq,87}-T^{2}_{eq,85}$ is thus accessible. $T^{2}_{eq,i}$ corresponding to the temporal part of the scale factor (\textit{cf.} Eq. (\ref{T eq}) in appendix \ref{Appendix Response function}) associated to the isotopes $i$. Figure \ref{difference in response functions} shows the behavior of this difference as a function of the Rabi frequency mismatch during the first and third pulses. The Rabi frequency mismatch is tuned by changing the micro-wave power at 3.035 GHz injected into the phase modulator, modifying both Raman pairs powers. The experimental value of the difference in response functions follows the same trend as the theoretical prediction derived from Eq. (\ref{scale factor}) which confirms the validity of the predicted scale factor.

Figure \ref{normalized transfer function} displays the acceleration transfer function of a differential dual-species atom interferometer with a wave-vector mismatch to which is added a Rabi frequency mismatch corresponding to our experimental conditions (differential 1 curve). It is very interesting to notice that in addition to the rejection loss at low frequency, the interferometer cut off becomes also less efficient at higher frequency. In our experiment, the Rabi frequency mismatch mainly comes from the impact of additional laser lines as it has already been explained (\textit{cf.} section \ref{Previous results and improvements} and appendix \ref{Appendix additional laser lines}). In these conditions the vibration rejection ratio limitation is evaluated at 100 dB.

\subsection{Limit - Drop of contrast because of inertial effects}
\label{Drop of contrast because of inertial effects}

\begin{figure}
\centerline{\includegraphics[width=8cm]{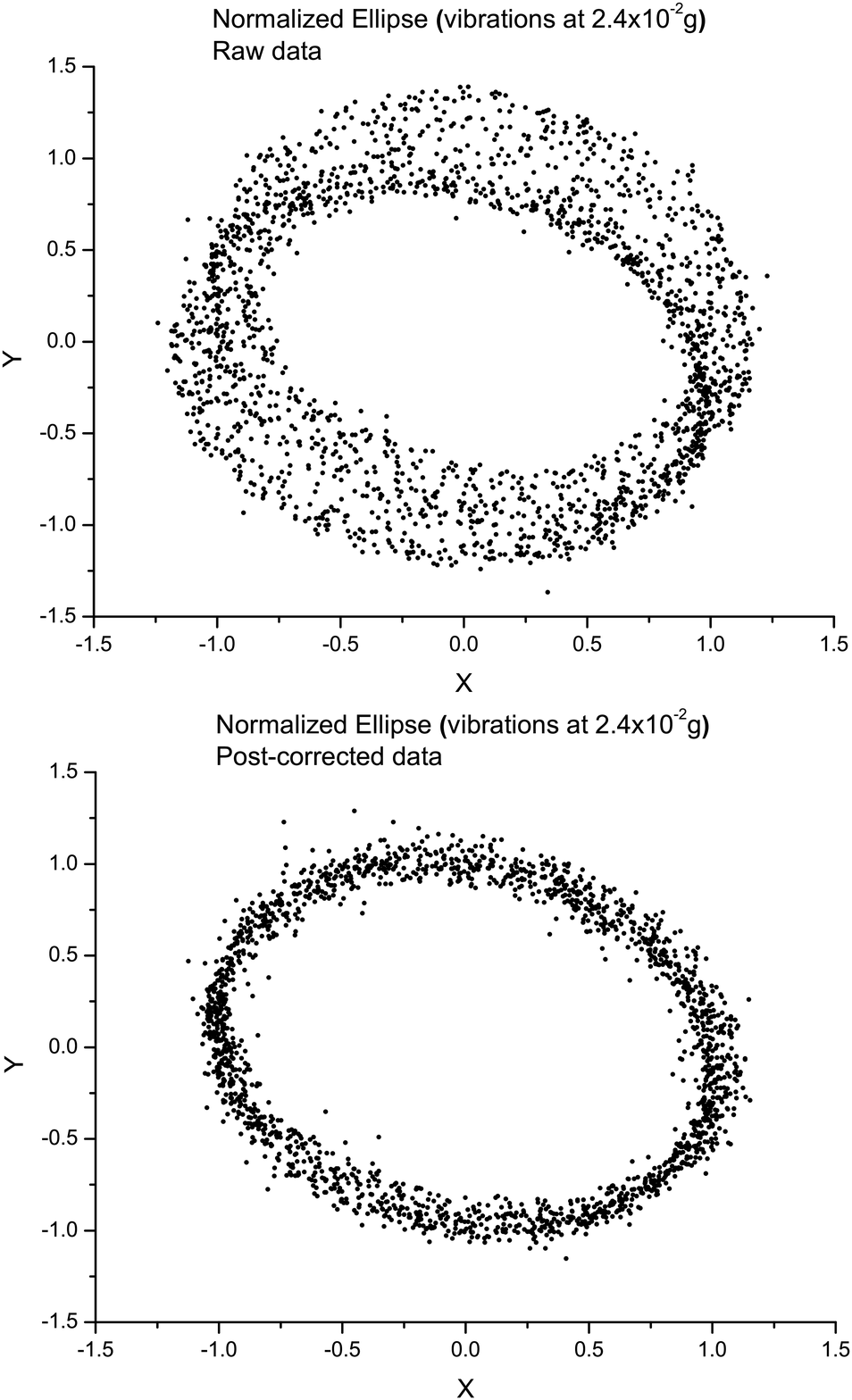}}
\caption
{
Normalized ellipses obtained with a vibration level of $2.4\times10^{-2} g$. At the top, the ellipse is plotted with the raw data. At the bottom, the ellipse with a better SNR resulting from the post-correction of spurious additional inertial effects. This correction is achievable thanks to the hybridization of the atom interferometer with classical inertial sensors.
}
\label{post corrected ellipse}
\end{figure}

During the vibration rejection experiment, additional accelerations and rotations emerge. These inertial effects will induce perturbations of the population offsets and a drop of the interferometer contrast. They can be visualized in Fig. \ref{3 ellipses} where the larger the vibrations are the noisier the ellipse is. It results a drop of the SNR and thus a deterioration of the resolution of the sensor. Nevertheless, if the differential atom accelerometer is hybridized with others inertial sensors, these spurious inertial effects can be measured and post-corrected. The Figure \ref{post corrected ellipse} shows the increase of the SNR after this post-correction. This is what has been done to obtain the vibration rejection ratio of 94 dB (cf. Fig. \ref{vibration rejection}).

At vibration amplitudes greater than $\sim$ 1 m$g$, the accelerations in the direction of $\vec{k}_{\scriptstyle\textrm{eff}}$ are large enough that the Raman pulses are Doppler shifted out of the Raman resonance condition. This leads to an estimated drop of contrast of 20 \% considering vibrations level of $\sim$ 40 m$g$ (peak to peak ; in the following, all the numbers will be given according to this level of vibrations).

Then, these vertical accelerations displace the atoms from the ideal detection position by a distance up to 1.2 mm. The detection efficiency is thus modified in a different way for each detection pulse. The normalization of atomic populations is also modified which creates fluctuations of the ellipse center up to 5 \%.

The interferometric phase depends on the transverse motion of atoms \cite{Louchet2011}. The rotations can thus induce a drop of contrast because of the averaging over the spatial and the velocity spread distributions of the atomic clouds. The rotations induce a drop of contrast through Coriolis effect \cite{Lan2012}, estimated at almost 90 \%, for a temperature of the atomic clouds of 2 $\mu$K, which is the most disruptive inertial effect. Moreover, angular accelerations, respectively centrifugal accelerations, may create additional phase shifts, as a function of the position of the atoms, which are averaged over the spatial distribution (of about 3 mm FWHM). The resulting drop of contrast is estimated at about 60 \%, respectively negligible.

\begin{figure}
\centerline{\includegraphics[width=10cm]{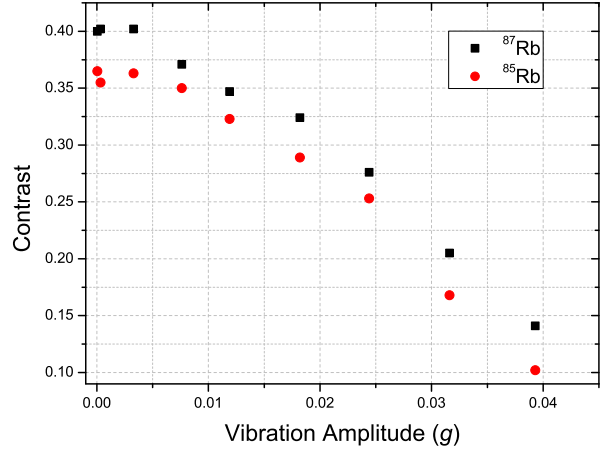}}
\caption
{
Drop of contrast as a function of the amplitude of vibrations for $^{87}$Rb (black squares) and for $^{85}$Rb (red circles). The drop being induced by spurious inertial effects.
}
\label{contrast VS vibrations}
\end{figure}

Figure \ref{contrast VS vibrations} presents the experimental drop of contrast induced by all spurious inertial effects previously mentioned for both isotopes of rubidium. At low vibration amplitude, the $^{85}$Rb contrast is lower than the $^{87}$Rb contrast. This is explained by the larger impact of additional laser lines on $^{85}$Rb and by its higher temperature. The same relation between the contrast is also observed at higher vibration level. This is also mainly explained by the higher temperature of $^{85}$Rb making the drop of contrast due to the Coriolis effect and angular accelerations stronger than for $^{87}$Rb.

In conclusion on the vibration noise rejection and its limitations, the rejection ratio of 94 dB (corresponding to a factor 50 000) experimentally measured is close to the theoretical limits estimated at 100 dB. This demonstrates that our differential atom interferometer works very close to its ultimate performances concerning the common-mode vibration noise rejection. The remaining difference is attributed to the non-perfect correction of the additional inertial effects induced by the excitation.

\section{Conclusion}
\label{Conclusion}

This experiment is intended to experimentally study some limits which could affect a quantum based test of the WEP by atom interferometry.

In this paper we have focused on the sensitivity and the resolution of our dual-species atom interferometer for measuring the differential acceleration and the profits on the common-mode vibration noise rejection. We did not study the accuracy of the measurement by estimating the systematic effects. Nevertheless, the theoretical study about the differential phase (\textit{cf.} Table \ref{differential phase table}) allows to state about the intrinsic phase shift inherent to the differential measurement by atom interferometry.

The free-fall accelerations of both test bodies are simultaneously compared and leads to a WEP test by extracting the E\"{o}tv\"{o}s ratio
\begin{equation}\label{Eotvos ratio} 
\eta_{^{87}\textrm{Rb},^{85}\textrm{Rb}} \equiv 2\frac{a_{87}-a_{85}}{(a_{87}+a_{85})} = 2\frac{\Delta a}{(a_{87}+a_{85})}
\end{equation}
The WEP is violated when $\eta_{^{87}\textrm{Rb},^{85}\textrm{Rb}}\neq0$ that is why the differential acceleration $\Delta a$ needs to be measured in an extremely accurate and sensitive way.

Concerning the accuracy, the WEP violation signal, mainly the term 1 in Table \ref{differential phase table}, is aimed to be detected. All additional phase shifts larger than this signal, coming from the mutual constant acceleration $a$ because of different scale factors between both isotopes (terms 5 to 7), from the atom velocity (terms 11 to 14), from experimental systematics effects (term 15, as light shifts, Zeeman effects, wave-front aberrations, gravity gradient, self-gravity, magnetic fields, mean field effects ...) must be accurately known and/or estimated to be canceled.

Concerning atom velocities, when Rabi frequencies are not equal between the first and the last light pulses (for instance because of the cloud expansion in a given longitudinal laser intensity profile) the dual-species interferometer is sensitive to the velocity difference between both species (term 11 to 14 and mainly term 11). By naming $\delta\Omega$ the Rabi frequency mismatch between the light pulses, the condition Term 11 $<$ Term 1 leads to the following condition on the velocity difference between both isotopes : $\Delta v < \frac{\Delta a T^{2} \Omega}{\delta\Omega/\Omega}$. By considering $\Delta a = 10^{-14}$ m.s$^{-2}$, $T=1$ s, $\tau = 100~\mu$s, $\delta\Omega/\Omega=10^{-2}$ we obtain $\Delta v < 2\times10^{-8}$ m.s$^{-1}$. This condition is experimentally challenging but remains below the conditions on the differential velocity targeted by STE-QUEST to counteract effects as the gravity gradient or the Coriolis acceleration \cite{Aguilera2014}.

Concerning the sensitivity, future experiments dedicated to test the WEP aim to be limited by the quantum projection noise during the atomic populations detection \cite{Yver-Leduc2003,Doring2010}. Such working regimes are currently reached in state of the art quantum inertial sensors. The main source of noise, which is inherently linked to the acceleration measurement, and which could thus impair the sensitivity is obviously the vibration noise, \textit{i.e.} $\tilde{a}$ terms in Table \ref{differential phase table}. As it has been previously explained, a simultaneous handling of both quantum test bodies is required to benefit from a common-mode vibration noise rejection. The use of both isotopes of the same species is here a strong advantage because it allows to take advantage of high rejection levels (as demonstrated in this paper) which ensure the high reliability of the differential accelerometer in a large range of environmental conditions. On ground, this allows to release some technological constraints on the isolating vibration systems for instance. In micro-gravity environments, according to terms 8 and 9, the use of $^{87}$Rb and $^{85}$Rb allows to work with white noise levels, in the interferometer bandwidth, up to $2\times10^{-4}$ m.s$^{-2}/\sqrt{Hz}$ by considering a relative difference of $10^{-9}$ on wave vectors and $10^{-3}$ on Rabi frequencies, or even up to $1\times10^{-3}$ m.s$^{-2}/\sqrt{Hz}$ if a 10$^{-4}$ level on the Rabi frequency match can be achieved (the values of $k$, $T$, $\tau$ has been here chosen according to STE-QUEST parameters \cite{Tino2013,Altschul2015,Aguilera2014}). This is a strong argument for the use of these two isotopes for testing the WEP in the international space station (ISS) or in a satellite. Indeed the vibration noise in the ISS is quite high because of human activity, whereas in a satellite, the vibration levels remain little-known and strongly depend on the satellite itself and on all the instruments that are loaded in there. The use of $^{87}$Rb and $^{85}$Rb thus offers a strong safety in order to counteract a large number of spurious and unexpected environmental effects.


We have reported here the realization and the characterization of a simultaneous dual-species atom interferometer which handles both $^{87}$Rb and $^{85}$Rb. We carefully studied the expression of the differential phase and the measurement scale factor, thanks to the sensitivity response function formalism, and explained the method used to derive the differential phase. After having tackled the main limitations of our previous device (the impact of additional laser lines and the non-linearities in the detection scheme) we showed that our differential accelerometer currently reaches a sensitivity of $1.23\times10^{-7}g$ at 1 s and a resolution of $2\times10^{-9}g$ after an integration time of few hours only. These results have been obtained with a small and compact atom sensor. The simultaneous differential measurement allowed us to exhibit a vibration rejection factor of 50 000 which proves the reliability of such a sensor in a large range of environmental conditions and its utility for future projects aiming to test the WEP with matter-waves. A next step to improve the current results could be to implement a new detection scheme for better counting the atomic populations of both species. Finally, it is necessary to work further on the correction of the systematic effects related to the measurement in order to reach an accuracy on $\eta$ at a level of $10^{-9}$. 

\begin{acknowledgments}

We thanks the French Defense Agency (DGA) for the financial support allowing the realization of the experimental setup.

\end{acknowledgments}

\appendix


\section{Response function of a Mach-Zehnder type atom interferometer to acceleration}
\label{Appendix Response function}

Initially developed for atomic clocks \cite{Dick1987}, the sensitivity function formalism is well adapted to evaluate the response of the interferometer to laser phase fluctuations and thus to acceleration.

Considering an infinitesimal phase step $\delta\phi$ of the Raman laser phase occurring at time $t$ the sensitivity function is defined as in Eq.(\ref{sensitivity function definition}). We make the same assumptions than in \cite{Cheinet2008}, \textit{i.e.} the laser waves are considered as pure plane waves, the Rabi frequencies are constant during a pulse (square pulses) and the resonance condition is fulfilled. Moreover we have extended this framework by assuming that the Rabi frequencies are not necessarily equal between each pulse ($\Omega_{1}\neq\Omega_{2}\neq\Omega_{3}$). We followed the same matrix calculation approach than in \cite{Cheinet2008}. With a time origin in the middle of the interferometer, the expression of the sensitivity function is given by 

\begin{equation}\label{sensitivity function}
g_{s}(t) =
\begin{cases}
0, & \text{$t \leq -T-2\tau$} \vspace{4pt}\\
-\frac{\sin\left[\Omega_{1}(t+T+2\tau)\right]}{\sin(\Omega_{1}\tau)}, & \text{$-T-2\tau \leq t \leq -T-\tau$} \vspace{4pt}\\
-1, & \text{$-T-\tau \leq t \leq -\tau$} \vspace{4pt}\\
\frac{\sin(\Omega_{2}t)}{\sin(\Omega_{2}\tau)}, & \text{$-\tau \leq t \leq \tau$} \vspace{4pt}\\
1, & \text{$\tau \leq t \leq T+\tau$} \vspace{4pt}\\
-\frac{\sin\left[\Omega_{3}(t-T-2\tau)\right]}{\sin(\Omega_{3}\tau)}, & \text{$T+\tau \leq t \leq T+2\tau$} \vspace{4pt}\\
0, & \text{$t \geq T+2\tau$} \\ 
\end{cases}
\end{equation}

The interferometric phase $\Delta\Phi$ can then be derived by the formula 
\begin{equation}\label{interferometric phase}
\Delta\Phi = \int_{-\infty}^{+\infty} g_{s}(t)\frac{\text{d}\varphi(t)}{\text{d}t}\,\text{d}t
\end{equation}
where $\varphi(t)$ is the phase of the Raman laser seen by the atoms. In the frame associated to the free falling atom, this phase is equal to $\varphi(t)=\vec{k}_{\scriptstyle\textrm{eff}}\cdot\vec{r}(t)$, with $\vec{r}(t)$ being the position of the Raman mirror compared to the atom. The sensitivity function corresponds to the response of the interferometer to the velocity of the free falling atom. Integrating by parts the Eq.(\ref{interferometric phase}), it can be shown that the interferometric phase can be expressed as in Eq.(\ref{phase with response functions}). The response function of the interferometer to acceleration is thus the primitive integral of the sensitivity function (\textit{cf}. Eq.(\ref{response function})) whose complete expression is 

\begin{widetext}
\begin{equation}\label{response function2}
f(t) =
\begin{cases}
\frac{1}{\Omega_{3}}\tan\left(\frac{\Omega_{3}\tau}{2}\right)-\frac{1}{\Omega_{1}}\tan\left(\frac{\Omega_{1}\tau}{2}\right), & \text{$t \leq -T-2\tau$} \vspace{4pt}\\
\frac{1}{\Omega_{1}\sin(\Omega_{1}\tau)}\left(\cos(\Omega_{1}\tau)-\cos\left[\Omega_{1}(t+T+2\tau)\right]\right)+\frac{1}{\Omega_{3}}\tan\left(\frac{\Omega_{3}\tau}{2}\right), & \text{$-T-2\tau \leq t \leq -T-\tau$} \vspace{4pt}\\
t+T+\tau+\frac{1}{\Omega_{3}}\tan\left(\frac{\Omega_{3}\tau}{2}\right), & \text{$-T-\tau \leq t \leq -\tau$} \vspace{4pt}\\
\frac{1}{\Omega_{2}\sin(\Omega_{2}\tau)}\left(\cos(\Omega_{2}t)-\cos(\Omega_{2}\tau)\right)+T+\frac{1}{\Omega_{3}}\tan\left(\frac{\Omega_{3}\tau}{2}\right), & \text{$-\tau \leq t \leq \tau$} \vspace{4pt}\\
-t+T+\tau+\frac{1}{\Omega_{3}}\tan\left(\frac{\Omega_{3}\tau}{2}\right), & \text{$\tau \leq t \leq T+\tau$} \vspace{4pt}\\
\frac{1}{\Omega_{3}\sin(\Omega_{3}\tau)}\left(1-\cos\left[\Omega_{3}(t-T-2\tau)\right]\right), & \text{$T+\tau \leq t \leq T+2\tau$} \vspace{4pt}\\
0, & \text{$t \geq T+2\tau$} \\ 
\end{cases}
\end{equation}
\end{widetext}

This triangle-shaped function (\textit{cf.} FIG.\ref{time response function}) evolves as $\cos(\Omega_{i}t)$ during the pulse $i$ and is maximum in the middle of the interferometer where the separation between both arms of the interferometer is the largest. The parity break of the interferometer symmetry between the first and the third Raman pulse makes it sensitive to the atom velocity. That is why $f(t)$ is not perfectly zero for $t \leq -T-2\tau$.

Thanks to this formalism, the expression of the scale factor $S$ impacting the measurement of a constant acceleration can be estimated by taken into account the finite duration of the Raman pulses :

\begin{equation}\label{scale factor 2}
\begin{array}{lll}
S&=& k_{\mathrm{eff}} \int_{-\infty}^{+\infty}g_{s}\left(t\right)t\,dt \vspace{4pt}\\

&=& k_{\mathrm{eff}} \Big[\Big. \vspace{4pt}\\

&& \left( T+2\tau \right)  \left( T+\frac{1}{\Omega_{1}}\tan\left(\frac{\Omega_{1}\tau}{2}\right) +\frac{1}{\Omega_{3}}\tan\left(\frac{\Omega_{3}\tau}{2}\right) \right) \vspace{4pt}\\

&& +\left(\frac{2}{\Omega_{2}^{2}}-\frac{1}{\Omega_{1}^{2}}-\frac{1}{\Omega_{3}^{2}}\right)  \vspace{4pt}\\

&& \left.-\tau\left( 2\frac{\mathrm{cotan}\left(\Omega_{2}\tau\right)}{\Omega_{2}}-\frac{\mathrm{cotan}\left(\Omega_{1}\tau\right)}{\Omega_{1}}-\frac{\mathrm{cotan}\left(\Omega_{3}\tau\right)}{\Omega_{3}} \right) \right]

\end{array}
\end{equation}

The temporal part of the scale factor $T^{2}_{eq}$ is thus given at first order by :

\begin{equation}\label{T eq}
T^{2}_{eq}=\left( T+2\tau \right)  \left( T+\frac{1}{\Omega_{1}}\tan\left(\frac{\Omega_{1}\tau}{2}\right) +\frac{1}{\Omega_{3}}\tan\left(\frac{\Omega_{3}\tau}{2}\right) \right)
\end{equation}


\section{Impact of additional laser lines generated by modulation on the atom interferometer}
\label{Appendix additional laser lines}

See also reference \cite{Carraz2012} for a more comprehensive approach.

Let us consider an atom corresponding to a $\Lambda$-type three-level system with two ground states $|a\rangle$ and $|b\rangle$ separated by an energy $\hbar\omega_{G}$ and an excited state $|e\rangle$ separated by $\hbar\omega_{0}$ from state $|b\rangle$. The atom interacts with a laser which is retro-reflected by a mirror at a position $z_{M}$ from the atom. The laser spectrum is composed of laser lines separated in frequency by $\omega_{G}$ and centered around the laser carrier pulsation $\omega_{L}$. This kind of spectrum is obtained with a phase modulated laser at frequency $\omega_{G}$ ($cf.$ spectrum corresponding to one species in Fig.\ref{Raman laser spectrum}). The electric field experienced by the atom can be decomposed on the $e^{in\omega_{G}t}$ basis thanks to the Bessel functions $J_{n}$, and written as the sum of
\begin{equation}\label{electric field}
\begin{array}{ll}
E_{\scriptstyle\textrm{down}}=E_{0}e^{i\omega_{L} t}\sum_{n=-\infty}^{+\infty}{i^{n}J_{n}(\phi)e^{in\omega_{G}t}} \vspace{0.25cm}+c.c \\
E_{\scriptstyle\textrm{up}}=E_{0}e^{i\omega_{L}(t-\frac{2z_{M}}{c})}\sum_{n=-\infty}^{+\infty}{i^{n}J_{n}(\phi)e^{in\omega_{G}(t-\frac{2z_{M}}{c})}}+c.c
\end{array}
\end{equation}
where $E_{\scriptstyle\textrm{down}}$ and $E_{\scriptstyle\textrm{up}}$ corresponds to the electric field respectively going downward and upward and $\phi$ to the depth of modulation.

The next step is to describe the probability amplitude to go from state $|a\rangle$ to state $|b\rangle$ when this electric field is turned on. We consider only a two-photon Raman transition with counter-propagating beams (the others co-propagating and opposite counter-propagating being neglected because of the Doppler shift induced by the free fall of atoms). Each component $E(n)$ of the electric field couples the two states with an associated Rabi frequency $\Omega_{n}$ proportional to
\begin{equation}\label{Raman Rabi frequency}
\Omega_{n}\propto\frac{E_{\scriptstyle\textrm{down}}^{*}(n+1)E_{\scriptstyle\textrm{up}}(n)}{\Delta-n\omega_{G}}
\end{equation}
where $\Delta=\omega_{L}-\omega_{G}$. The resulting ``overall" two-photon Rabi frequency $\Omega_{|a\rangle\rightarrow|b\rangle}$ is thus given by
\begin{equation}\label{overall Raman Rabi frequency}
\Omega_{|a\rangle\rightarrow|b\rangle}=\sum_{n=-\infty}^{+\infty}{\Omega_{n}}
\end{equation}
In this formalism we do not consider the external state of the atom, the degeneracy of the quantum states due to the slight difference in momentum $\hbar\Delta k=2\hbar\omega_{G}/c$ between the different Raman pairs is neglected. This approximation is a bit rough but allows to simply explain the problem.The overall Rabi frequency is thus expressed as 
\begin{equation}\label{overall Raman Rabi frequency 2}
\Omega_{|a\rangle\rightarrow|b\rangle} \propto \sum_{n=-\infty}^{+\infty} \left[ \frac{J_{n}(\phi)J_{n+1}(\phi)}{\Delta-n\omega_{G}}e^{in\Delta k z_{M}} + c.c\right]
\end{equation}
The terms in this sum interfere, which has the effect of spatially modulate the Rabi frequency. The global interferometer contrast will depend on the distance between the atoms and the mirror ($z_{M}$) and thus also on the time when each Raman pulse occurs. 


\bibliography{bibliography_of_article2}

\begin{thebibliography}{59}%
\makeatletter
\providecommand \@ifxundefined [1]{%
 \@ifx{#1\undefined}
}%
\providecommand \@ifnum [1]{%
 \ifnum #1\expandafter \@firstoftwo
 \else \expandafter \@secondoftwo
 \fi
}%
\providecommand \@ifx [1]{%
 \ifx #1\expandafter \@firstoftwo
 \else \expandafter \@secondoftwo
 \fi
}%
\providecommand \natexlab [1]{#1}%
\providecommand \enquote  [1]{``#1''}%
\providecommand \bibnamefont  [1]{#1}%
\providecommand \bibfnamefont [1]{#1}%
\providecommand \citenamefont [1]{#1}%
\providecommand \href@noop [0]{\@secondoftwo}%
\providecommand \href [0]{\begingroup \@sanitize@url \@href}%
\providecommand \@href[1]{\@@startlink{#1}\@@href}%
\providecommand \@@href[1]{\endgroup#1\@@endlink}%
\providecommand \@sanitize@url [0]{\catcode `\\12\catcode `\$12\catcode
  `\&12\catcode `\#12\catcode `\^12\catcode `\_12\catcode `\%12\relax}%
\providecommand \@@startlink[1]{}%
\providecommand \@@endlink[0]{}%
\providecommand \url  [0]{\begingroup\@sanitize@url \@url }%
\providecommand \@url [1]{\endgroup\@href {#1}{\urlprefix }}%
\providecommand \urlprefix  [0]{URL }%
\providecommand \Eprint [0]{\href }%
\providecommand \doibase [0]{http://dx.doi.org/}%
\providecommand \selectlanguage [0]{\@gobble}%
\providecommand \bibinfo  [0]{\@secondoftwo}%
\providecommand \bibfield  [0]{\@secondoftwo}%
\providecommand \translation [1]{[#1]}%
\providecommand \BibitemOpen [0]{}%
\providecommand \bibitemStop [0]{}%
\providecommand \bibitemNoStop [0]{.\EOS\space}%
\providecommand \EOS [0]{\spacefactor3000\relax}%
\providecommand \BibitemShut  [1]{\csname bibitem#1\endcsname}%
\let\auto@bib@innerbib\@empty
\bibitem [{\citenamefont {Bord\'{e}}(1989)}]{Borde1989}%
  \BibitemOpen
  \bibfield  {author} {\bibinfo {author} {\bibfnamefont {C.~J.}\ \bibnamefont
  {Bord\'{e}}},\ }\href@noop {} {\bibfield  {journal} {\bibinfo  {journal}
  {Phys. Lett. A}\ }\textbf {\bibinfo {volume} {140}},\ \bibinfo {pages} {10}
  (\bibinfo {year} {1989})}\BibitemShut {NoStop}%
\bibitem [{\citenamefont {Tino}\ and\ \citenamefont
  {Kasevich}(2014)}]{Tino2014}%
  \BibitemOpen
  \bibfield  {author} {\bibinfo {author} {\bibfnamefont {G.~M.}\ \bibnamefont
  {Tino}}\ and\ \bibinfo {author} {\bibfnamefont {M.~A.}\ \bibnamefont
  {Kasevich}},\ }\href@noop {} {\emph {\bibinfo {title} {Atom Interferometry,
  Proceedings of the International School of Physics “ Enrico Fermi”,
  Course CLXXXVIII}}}\ (\bibinfo  {publisher} {IOS Press, Amsterdam},\ \bibinfo
  {year} {2014})\BibitemShut {NoStop}%
\bibitem [{\citenamefont {Peters}\ \emph {et~al.}(2001)\citenamefont {Peters},
  \citenamefont {Chung},\ and\ \citenamefont {Chu}}]{Peters2001}%
  \BibitemOpen
  \bibfield  {author} {\bibinfo {author} {\bibfnamefont {A.}~\bibnamefont
  {Peters}}, \bibinfo {author} {\bibfnamefont {K.~Y.}\ \bibnamefont {Chung}}, \
  and\ \bibinfo {author} {\bibfnamefont {S.}~\bibnamefont {Chu}},\ }\href@noop
  {} {\bibfield  {journal} {\bibinfo  {journal} {Metrologia}\ }\textbf
  {\bibinfo {volume} {38}},\ \bibinfo {pages} {25} (\bibinfo {year}
  {2001})}\BibitemShut {NoStop}%
\bibitem [{\citenamefont {Gillot}\ \emph {et~al.}(2014)\citenamefont {Gillot},
  \citenamefont {Francis}, \citenamefont {Landragin}, \citenamefont {Pereira
  Dos~Santos},\ and\ \citenamefont {Merlet}}]{Gillot2014}%
  \BibitemOpen
  \bibfield  {author} {\bibinfo {author} {\bibfnamefont {P.}~\bibnamefont
  {Gillot}}, \bibinfo {author} {\bibfnamefont {O.}~\bibnamefont {Francis}},
  \bibinfo {author} {\bibfnamefont {A.}~\bibnamefont {Landragin}}, \bibinfo
  {author} {\bibfnamefont {F.}~\bibnamefont {Pereira Dos~Santos}}, \ and\
  \bibinfo {author} {\bibfnamefont {S.}~\bibnamefont {Merlet}},\ }\href@noop {}
  {\bibfield  {journal} {\bibinfo  {journal} {Metrologia}\ }\textbf {\bibinfo
  {volume} {51}},\ \bibinfo {pages} {15} (\bibinfo {year} {2014})}\BibitemShut
  {NoStop}%
\bibitem [{\citenamefont {Hu}\ \emph {et~al.}(2013)\citenamefont {Hu},
  \citenamefont {Sun}, \citenamefont {Duan}, \citenamefont {Zhou},
  \citenamefont {Chen}, \citenamefont {Zhan}, \citenamefont {Zhang},\ and\
  \citenamefont {Luo}}]{Hu2013}%
  \BibitemOpen
  \bibfield  {author} {\bibinfo {author} {\bibfnamefont {Z.~K.}\ \bibnamefont
  {Hu}}, \bibinfo {author} {\bibfnamefont {B.~L.}\ \bibnamefont {Sun}},
  \bibinfo {author} {\bibfnamefont {X.~C.}\ \bibnamefont {Duan}}, \bibinfo
  {author} {\bibfnamefont {M.~K.}\ \bibnamefont {Zhou}}, \bibinfo {author}
  {\bibfnamefont {L.~L.}\ \bibnamefont {Chen}}, \bibinfo {author}
  {\bibfnamefont {S.}~\bibnamefont {Zhan}}, \bibinfo {author} {\bibfnamefont
  {Q.~Z.}\ \bibnamefont {Zhang}}, \ and\ \bibinfo {author} {\bibfnamefont
  {J.}~\bibnamefont {Luo}},\ }\href@noop {} {\bibfield  {journal} {\bibinfo
  {journal} {Physical Review A}\ }\textbf {\bibinfo {volume} {88}},\ \bibinfo
  {pages} {043610} (\bibinfo {year} {2013})}\BibitemShut {NoStop}%
\bibitem [{\citenamefont {Bidel}\ \emph {et~al.}(2013)\citenamefont {Bidel},
  \citenamefont {Carraz}, \citenamefont {Charri\`{e}re}, \citenamefont
  {Cadoret}, \citenamefont {Zahzam},\ and\ \citenamefont
  {Bresson}}]{Bidel2013}%
  \BibitemOpen
  \bibfield  {author} {\bibinfo {author} {\bibfnamefont {Y.}~\bibnamefont
  {Bidel}}, \bibinfo {author} {\bibfnamefont {O.}~\bibnamefont {Carraz}},
  \bibinfo {author} {\bibfnamefont {R.}~\bibnamefont {Charri\`{e}re}}, \bibinfo
  {author} {\bibfnamefont {M.}~\bibnamefont {Cadoret}}, \bibinfo {author}
  {\bibfnamefont {N.}~\bibnamefont {Zahzam}}, \ and\ \bibinfo {author}
  {\bibfnamefont {A.}~\bibnamefont {Bresson}},\ }\href@noop {} {\bibfield
  {journal} {\bibinfo  {journal} {Appl. Phys. Lett.}\ }\textbf {\bibinfo
  {volume} {102}},\ \bibinfo {pages} {144107} (\bibinfo {year}
  {2013})}\BibitemShut {NoStop}%
\bibitem [{\citenamefont {Hauth}\ \emph {et~al.}(2013)\citenamefont {Hauth},
  \citenamefont {Freier}, \citenamefont {Schkolnik}, \citenamefont {Senger},
  \citenamefont {Schmidt},\ and\ \citenamefont {Peters}}]{Hauth2013}%
  \BibitemOpen
  \bibfield  {author} {\bibinfo {author} {\bibfnamefont {M.}~\bibnamefont
  {Hauth}}, \bibinfo {author} {\bibfnamefont {C.}~\bibnamefont {Freier}},
  \bibinfo {author} {\bibfnamefont {V.}~\bibnamefont {Schkolnik}}, \bibinfo
  {author} {\bibfnamefont {A.}~\bibnamefont {Senger}}, \bibinfo {author}
  {\bibfnamefont {M.}~\bibnamefont {Schmidt}}, \ and\ \bibinfo {author}
  {\bibfnamefont {A.}~\bibnamefont {Peters}},\ }\href@noop {} {\bibfield
  {journal} {\bibinfo  {journal} {Applied Physics B}\ }\textbf {\bibinfo
  {volume} {113}},\ \bibinfo {pages} {49} (\bibinfo {year} {2013})}\BibitemShut
  {NoStop}%
\bibitem [{\citenamefont {McGuirk}\ \emph {et~al.}(2002)\citenamefont
  {McGuirk}, \citenamefont {Foster}, \citenamefont {Fixler}, \citenamefont
  {Snadden},\ and\ \citenamefont {Kasevich}}]{McGuirk2002}%
  \BibitemOpen
  \bibfield  {author} {\bibinfo {author} {\bibfnamefont {J.~M.}\ \bibnamefont
  {McGuirk}}, \bibinfo {author} {\bibfnamefont {G.~T.}\ \bibnamefont {Foster}},
  \bibinfo {author} {\bibfnamefont {J.~B.}\ \bibnamefont {Fixler}}, \bibinfo
  {author} {\bibfnamefont {M.~J.}\ \bibnamefont {Snadden}}, \ and\ \bibinfo
  {author} {\bibfnamefont {M.~A.}\ \bibnamefont {Kasevich}},\ }\href@noop {}
  {\bibfield  {journal} {\bibinfo  {journal} {Phys. Rev. A}\ }\textbf {\bibinfo
  {volume} {65}},\ \bibinfo {pages} {033608} (\bibinfo {year}
  {2002})}\BibitemShut {NoStop}%
\bibitem [{\citenamefont {Sorrentino}\ \emph {et~al.}(2014)\citenamefont
  {Sorrentino}, \citenamefont {Bodart}, \citenamefont {Cacciapuoti},
  \citenamefont {Lien}, \citenamefont {Prevedelli}, \citenamefont {Rosi},
  \citenamefont {Salvi},\ and\ \citenamefont {Tino}}]{Sorrentino2014}%
  \BibitemOpen
  \bibfield  {author} {\bibinfo {author} {\bibfnamefont {F.}~\bibnamefont
  {Sorrentino}}, \bibinfo {author} {\bibfnamefont {Q.}~\bibnamefont {Bodart}},
  \bibinfo {author} {\bibfnamefont {L.}~\bibnamefont {Cacciapuoti}}, \bibinfo
  {author} {\bibfnamefont {Y.-H.}\ \bibnamefont {Lien}}, \bibinfo {author}
  {\bibfnamefont {M.}~\bibnamefont {Prevedelli}}, \bibinfo {author}
  {\bibfnamefont {G.}~\bibnamefont {Rosi}}, \bibinfo {author} {\bibfnamefont
  {L.}~\bibnamefont {Salvi}}, \ and\ \bibinfo {author} {\bibfnamefont {G.~M.}\
  \bibnamefont {Tino}},\ }\href@noop {} {\bibfield  {journal} {\bibinfo
  {journal} {Phys. Rev. A}\ }\textbf {\bibinfo {volume} {89}},\ \bibinfo
  {pages} {023607} (\bibinfo {year} {2014})}\BibitemShut {NoStop}%
\bibitem [{\citenamefont {Duan}\ \emph {et~al.}(2014)\citenamefont {Duan},
  \citenamefont {Zhou}, \citenamefont {Mao}, \citenamefont {Yao}, \citenamefont
  {Deng}, \citenamefont {Luo},\ and\ \citenamefont {Hu}}]{Duan2014}%
  \BibitemOpen
  \bibfield  {author} {\bibinfo {author} {\bibfnamefont {X.-C.}\ \bibnamefont
  {Duan}}, \bibinfo {author} {\bibfnamefont {M.-K.}\ \bibnamefont {Zhou}},
  \bibinfo {author} {\bibfnamefont {D.-K.}\ \bibnamefont {Mao}}, \bibinfo
  {author} {\bibfnamefont {H.-B.}\ \bibnamefont {Yao}}, \bibinfo {author}
  {\bibfnamefont {X.-B.}\ \bibnamefont {Deng}}, \bibinfo {author}
  {\bibfnamefont {J.}~\bibnamefont {Luo}}, \ and\ \bibinfo {author}
  {\bibfnamefont {Z.-K.}\ \bibnamefont {Hu}},\ }\href@noop {} {\bibfield
  {journal} {\bibinfo  {journal} {Phys. Rev. A}\ }\textbf {\bibinfo {volume}
  {90}},\ \bibinfo {pages} {023617} (\bibinfo {year} {2014})}\BibitemShut
  {NoStop}%
\bibitem [{\citenamefont {Gustavson}\ \emph {et~al.}(1997)\citenamefont
  {Gustavson}, \citenamefont {Bouyer},\ and\ \citenamefont
  {Kasevich}}]{Gustavson1997}%
  \BibitemOpen
  \bibfield  {author} {\bibinfo {author} {\bibfnamefont {T.~L.}\ \bibnamefont
  {Gustavson}}, \bibinfo {author} {\bibfnamefont {P.}~\bibnamefont {Bouyer}}, \
  and\ \bibinfo {author} {\bibfnamefont {M.~A.}\ \bibnamefont {Kasevich}},\
  }\href@noop {} {\bibfield  {journal} {\bibinfo  {journal} {Phys. Rev. Lett.}\
  }\textbf {\bibinfo {volume} {78}},\ \bibinfo {pages} {2046} (\bibinfo {year}
  {1997})}\BibitemShut {NoStop}%
\bibitem [{\citenamefont {Gauguet}\ \emph {et~al.}(2009)\citenamefont
  {Gauguet}, \citenamefont {Canuel}, \citenamefont {L\'ev\`eque}, \citenamefont
  {Chaibi},\ and\ \citenamefont {Landragin}}]{Gauguet2009}%
  \BibitemOpen
  \bibfield  {author} {\bibinfo {author} {\bibfnamefont {A.}~\bibnamefont
  {Gauguet}}, \bibinfo {author} {\bibfnamefont {B.}~\bibnamefont {Canuel}},
  \bibinfo {author} {\bibfnamefont {T.}~\bibnamefont {L\'ev\`eque}}, \bibinfo
  {author} {\bibfnamefont {W.}~\bibnamefont {Chaibi}}, \ and\ \bibinfo {author}
  {\bibfnamefont {A.}~\bibnamefont {Landragin}},\ }\href@noop {} {\bibfield
  {journal} {\bibinfo  {journal} {Phys. Rev. A}\ }\textbf {\bibinfo {volume}
  {80}},\ \bibinfo {pages} {063604} (\bibinfo {year} {2009})}\BibitemShut
  {NoStop}%
\bibitem [{\citenamefont {Tackmann}\ \emph {et~al.}(2012)\citenamefont
  {Tackmann}, \citenamefont {Berg}, \citenamefont {Schubert}, \citenamefont
  {Abend}, \citenamefont {Ertmer},\ and\ \citenamefont {Rasel}}]{Tackmann2012}%
  \BibitemOpen
  \bibfield  {author} {\bibinfo {author} {\bibfnamefont {G.}~\bibnamefont
  {Tackmann}}, \bibinfo {author} {\bibfnamefont {P.}~\bibnamefont {Berg}},
  \bibinfo {author} {\bibfnamefont {C.}~\bibnamefont {Schubert}}, \bibinfo
  {author} {\bibfnamefont {M.}~\bibnamefont {Abend}, \bibfnamefont
  {S.~andGilowski}}, \bibinfo {author} {\bibfnamefont {W.}~\bibnamefont
  {Ertmer}}, \ and\ \bibinfo {author} {\bibfnamefont {E.~M.}\ \bibnamefont
  {Rasel}},\ }\href@noop {} {\bibfield  {journal} {\bibinfo  {journal} {New
  Journal of Physics}\ }\textbf {\bibinfo {volume} {14}},\ \bibinfo {pages}
  {015002} (\bibinfo {year} {2012})}\BibitemShut {NoStop}%
\bibitem [{\citenamefont {Rosi}\ \emph {et~al.}(2015)\citenamefont {Rosi},
  \citenamefont {Cacciapuoti}, \citenamefont {Sorrentino}, \citenamefont
  {Menchetti}, \citenamefont {Prevedelli},\ and\ \citenamefont
  {Tino}}]{Rosi2015}%
  \BibitemOpen
  \bibfield  {author} {\bibinfo {author} {\bibfnamefont {G.}~\bibnamefont
  {Rosi}}, \bibinfo {author} {\bibfnamefont {L.}~\bibnamefont {Cacciapuoti}},
  \bibinfo {author} {\bibfnamefont {F.}~\bibnamefont {Sorrentino}}, \bibinfo
  {author} {\bibfnamefont {M.}~\bibnamefont {Menchetti}}, \bibinfo {author}
  {\bibfnamefont {M.}~\bibnamefont {Prevedelli}}, \ and\ \bibinfo {author}
  {\bibfnamefont {G.~M.}\ \bibnamefont {Tino}},\ }\href@noop {} {\bibfield
  {journal} {\bibinfo  {journal} {Phys. Rev. Lett.}\ }\textbf {\bibinfo
  {volume} {114}},\ \bibinfo {pages} {013001} (\bibinfo {year}
  {2015})}\BibitemShut {NoStop}%
\bibitem [{\citenamefont {Jeleki}(2005)}]{Jeleki2005}%
  \BibitemOpen
  \bibfield  {author} {\bibinfo {author} {\bibfnamefont {C.}~\bibnamefont
  {Jeleki}},\ }\href@noop {} {\bibfield  {journal} {\bibinfo  {journal}
  {Navigation}\ }\textbf {\bibinfo {volume} {2}},\ \bibinfo {pages} {1}
  (\bibinfo {year} {2005})}\BibitemShut {NoStop}%
\bibitem [{\citenamefont {Bouchendira}\ \emph {et~al.}(2011)\citenamefont
  {Bouchendira}, \citenamefont {Clad\'{e}}, \citenamefont
  {Guellati-Kh\'{e}lifa}, \citenamefont {Nez},\ and\ \citenamefont
  {Biraben}}]{Bouchendira2011}%
  \BibitemOpen
  \bibfield  {author} {\bibinfo {author} {\bibfnamefont {R.}~\bibnamefont
  {Bouchendira}}, \bibinfo {author} {\bibfnamefont {P.}~\bibnamefont
  {Clad\'{e}}}, \bibinfo {author} {\bibfnamefont {S.}~\bibnamefont
  {Guellati-Kh\'{e}lifa}}, \bibinfo {author} {\bibfnamefont {F.}~\bibnamefont
  {Nez}}, \ and\ \bibinfo {author} {\bibfnamefont {F.}~\bibnamefont
  {Biraben}},\ }\href@noop {} {\bibfield  {journal} {\bibinfo  {journal} {Phys.
  Rev. Lett.}\ }\textbf {\bibinfo {volume} {106}},\ \bibinfo {pages} {080801}
  (\bibinfo {year} {2011})}\BibitemShut {NoStop}%
\bibitem [{\citenamefont {Fixler}\ \emph {et~al.}(2007)\citenamefont {Fixler},
  \citenamefont {Foster}, \citenamefont {McGuirk},\ and\ \citenamefont
  {Kasevich}}]{Fixler2007}%
  \BibitemOpen
  \bibfield  {author} {\bibinfo {author} {\bibfnamefont {J.~B.}\ \bibnamefont
  {Fixler}}, \bibinfo {author} {\bibfnamefont {G.~T.}\ \bibnamefont {Foster}},
  \bibinfo {author} {\bibfnamefont {J.~M.}\ \bibnamefont {McGuirk}}, \ and\
  \bibinfo {author} {\bibfnamefont {M.~A.}\ \bibnamefont {Kasevich}},\
  }\href@noop {} {\bibfield  {journal} {\bibinfo  {journal} {Science}\ }\textbf
  {\bibinfo {volume} {315}},\ \bibinfo {pages} {74} (\bibinfo {year}
  {2007})}\BibitemShut {NoStop}%
\bibitem [{\citenamefont {Rosi}\ \emph {et~al.}(2014)\citenamefont {Rosi},
  \citenamefont {Sorrentino}, \citenamefont {Cacciapuoti}, \citenamefont
  {Prevedelli},\ and\ \citenamefont {Tino}}]{Rosi2014}%
  \BibitemOpen
  \bibfield  {author} {\bibinfo {author} {\bibfnamefont {G.}~\bibnamefont
  {Rosi}}, \bibinfo {author} {\bibfnamefont {F.}~\bibnamefont {Sorrentino}},
  \bibinfo {author} {\bibfnamefont {L.}~\bibnamefont {Cacciapuoti}}, \bibinfo
  {author} {\bibfnamefont {M.}~\bibnamefont {Prevedelli}}, \ and\ \bibinfo
  {author} {\bibfnamefont {G.~M.}\ \bibnamefont {Tino}},\ }\href@noop {}
  {\bibfield  {journal} {\bibinfo  {journal} {Nature}\ }\textbf {\bibinfo
  {volume} {510}},\ \bibinfo {pages} {518} (\bibinfo {year}
  {2014})}\BibitemShut {NoStop}%
\bibitem [{\citenamefont {Dimopoulos}\ \emph {et~al.}(2008)\citenamefont
  {Dimopoulos}, \citenamefont {Graham}, \citenamefont {Hogan}, \citenamefont
  {Kasevich},\ and\ \citenamefont {Rajendran}}]{Dimopoulos2008}%
  \BibitemOpen
  \bibfield  {author} {\bibinfo {author} {\bibfnamefont {S.}~\bibnamefont
  {Dimopoulos}}, \bibinfo {author} {\bibfnamefont {P.~W.}\ \bibnamefont
  {Graham}}, \bibinfo {author} {\bibfnamefont {J.~M.}\ \bibnamefont {Hogan}},
  \bibinfo {author} {\bibfnamefont {M.~A.}\ \bibnamefont {Kasevich}}, \ and\
  \bibinfo {author} {\bibfnamefont {S.}~\bibnamefont {Rajendran}},\ }\href@noop
  {} {\bibfield  {journal} {\bibinfo  {journal} {Phys. Rev. D}\ }\textbf
  {\bibinfo {volume} {78}},\ \bibinfo {pages} {122002} (\bibinfo {year}
  {2008})}\BibitemShut {NoStop}%
\bibitem [{\citenamefont {Ferrari}\ \emph {et~al.}(2006)\citenamefont
  {Ferrari}, \citenamefont {Poli}, \citenamefont {Sorrentino},\ and\
  \citenamefont {Tino}}]{Ferrari2006}%
  \BibitemOpen
  \bibfield  {author} {\bibinfo {author} {\bibfnamefont {G.}~\bibnamefont
  {Ferrari}}, \bibinfo {author} {\bibfnamefont {N.}~\bibnamefont {Poli}},
  \bibinfo {author} {\bibfnamefont {F.}~\bibnamefont {Sorrentino}}, \ and\
  \bibinfo {author} {\bibfnamefont {G.~M.}\ \bibnamefont {Tino}},\ }\href@noop
  {} {\bibfield  {journal} {\bibinfo  {journal} {Phys. Rev. Lett.}\ }\textbf
  {\bibinfo {volume} {97}},\ \bibinfo {pages} {060402} (\bibinfo {year}
  {2006})}\BibitemShut {NoStop}%
\bibitem [{\citenamefont {Wolf}\ \emph {et~al.}(2007)\citenamefont {Wolf},
  \citenamefont {Lemonde}, \citenamefont {Lambrecht}, \citenamefont {Bize},
  \citenamefont {Landragin},\ and\ \citenamefont {Clairon}}]{Wolf2007}%
  \BibitemOpen
  \bibfield  {author} {\bibinfo {author} {\bibfnamefont {P.}~\bibnamefont
  {Wolf}}, \bibinfo {author} {\bibfnamefont {P.}~\bibnamefont {Lemonde}},
  \bibinfo {author} {\bibfnamefont {A.}~\bibnamefont {Lambrecht}}, \bibinfo
  {author} {\bibfnamefont {S.}~\bibnamefont {Bize}}, \bibinfo {author}
  {\bibfnamefont {A.}~\bibnamefont {Landragin}}, \ and\ \bibinfo {author}
  {\bibfnamefont {A.}~\bibnamefont {Clairon}},\ }\href@noop {} {\bibfield
  {journal} {\bibinfo  {journal} {Phys. Rev. A}\ }\textbf {\bibinfo {volume}
  {75}},\ \bibinfo {pages} {063608} (\bibinfo {year} {2007})}\BibitemShut
  {NoStop}%
\bibitem [{\citenamefont {Fray}\ \emph {et~al.}(2004)\citenamefont {Fray},
  \citenamefont {Diez}, \citenamefont {H\"ansch},\ and\ \citenamefont
  {Weitz}}]{Fray2004}%
  \BibitemOpen
  \bibfield  {author} {\bibinfo {author} {\bibfnamefont {S.}~\bibnamefont
  {Fray}}, \bibinfo {author} {\bibfnamefont {C.~A.}\ \bibnamefont {Diez}},
  \bibinfo {author} {\bibfnamefont {T.~W.}\ \bibnamefont {H\"ansch}}, \ and\
  \bibinfo {author} {\bibfnamefont {M.}~\bibnamefont {Weitz}},\ }\href@noop {}
  {\bibfield  {journal} {\bibinfo  {journal} {Phys. Rev. Lett.}\ }\textbf
  {\bibinfo {volume} {93}},\ \bibinfo {pages} {240404} (\bibinfo {year}
  {2004})}\BibitemShut {NoStop}%
\bibitem [{\citenamefont {Bonnin}\ \emph {et~al.}(2013)\citenamefont {Bonnin},
  \citenamefont {Zahzam}, \citenamefont {Bidel},\ and\ \citenamefont
  {Bresson}}]{Bonnin2013}%
  \BibitemOpen
  \bibfield  {author} {\bibinfo {author} {\bibfnamefont {A.}~\bibnamefont
  {Bonnin}}, \bibinfo {author} {\bibfnamefont {N.}~\bibnamefont {Zahzam}},
  \bibinfo {author} {\bibfnamefont {Y.}~\bibnamefont {Bidel}}, \ and\ \bibinfo
  {author} {\bibfnamefont {A.}~\bibnamefont {Bresson}},\ }\href@noop {}
  {\bibfield  {journal} {\bibinfo  {journal} {Phys. Rew. A}\ }\textbf {\bibinfo
  {volume} {88}},\ \bibinfo {pages} {043615} (\bibinfo {year}
  {2013})}\BibitemShut {NoStop}%
\bibitem [{\citenamefont {Schlippert}\ \emph {et~al.}(2014)\citenamefont
  {Schlippert}, \citenamefont {Hartwig}, \citenamefont {Albers}, \citenamefont
  {Richardson}, \citenamefont {Schubert}, \citenamefont {Roura}, \citenamefont
  {Schleich}, \citenamefont {Ertmer},\ and\ \citenamefont
  {Rasel}}]{Schlippert2014}%
  \BibitemOpen
  \bibfield  {author} {\bibinfo {author} {\bibfnamefont {D.}~\bibnamefont
  {Schlippert}}, \bibinfo {author} {\bibfnamefont {J.}~\bibnamefont {Hartwig}},
  \bibinfo {author} {\bibfnamefont {H.}~\bibnamefont {Albers}}, \bibinfo
  {author} {\bibfnamefont {L.~L.}\ \bibnamefont {Richardson}}, \bibinfo
  {author} {\bibfnamefont {C.}~\bibnamefont {Schubert}}, \bibinfo {author}
  {\bibfnamefont {A.}~\bibnamefont {Roura}}, \bibinfo {author} {\bibfnamefont
  {W.~P.}\ \bibnamefont {Schleich}}, \bibinfo {author} {\bibfnamefont
  {W.}~\bibnamefont {Ertmer}}, \ and\ \bibinfo {author} {\bibfnamefont {E.~M.}\
  \bibnamefont {Rasel}},\ }\href@noop {} {\bibfield  {journal} {\bibinfo
  {journal} {Phys. Rev. Lett.}\ }\textbf {\bibinfo {volume} {112}},\ \bibinfo
  {pages} {203002} (\bibinfo {year} {2014})}\BibitemShut {NoStop}%
\bibitem [{\citenamefont {Tarallo}\ \emph {et~al.}(2014)\citenamefont
  {Tarallo}, \citenamefont {Mazzoni}, \citenamefont {Poli}, \citenamefont
  {Sutyrin}, \citenamefont {Zhang},\ and\ \citenamefont {Tino}}]{Tarallo2014}%
  \BibitemOpen
  \bibfield  {author} {\bibinfo {author} {\bibfnamefont {M.~G.}\ \bibnamefont
  {Tarallo}}, \bibinfo {author} {\bibfnamefont {T.}~\bibnamefont {Mazzoni}},
  \bibinfo {author} {\bibfnamefont {N.}~\bibnamefont {Poli}}, \bibinfo {author}
  {\bibfnamefont {D.~V.}\ \bibnamefont {Sutyrin}}, \bibinfo {author}
  {\bibfnamefont {X.}~\bibnamefont {Zhang}}, \ and\ \bibinfo {author}
  {\bibfnamefont {G.~M.}\ \bibnamefont {Tino}},\ }\href@noop {} {\bibfield
  {journal} {\bibinfo  {journal} {Phys. Rev. Lett.}\ }\textbf {\bibinfo
  {volume} {113}},\ \bibinfo {pages} {023005} (\bibinfo {year}
  {2014})}\BibitemShut {NoStop}%
\bibitem [{\citenamefont {Zhou}\ and\ \citenamefont {others
  ...}(2015)}]{Zhou2015}%
  \BibitemOpen
  \bibfield  {author} {\bibinfo {author} {\bibfnamefont {L.}~\bibnamefont
  {Zhou}}\ and\ \bibinfo {author} {\bibnamefont {others ...}},\ }\href@noop {}
  {\bibfield  {journal} {\bibinfo  {journal} {arXiv}\ }\textbf {\bibinfo
  {volume} {1503.004401v1}} (\bibinfo {year} {2015})}\BibitemShut {NoStop}%
\bibitem [{\citenamefont {Damour}(2012)}]{Damour2012}%
  \BibitemOpen
  \bibfield  {author} {\bibinfo {author} {\bibfnamefont {T.}~\bibnamefont
  {Damour}},\ }\href@noop {} {\bibfield  {journal} {\bibinfo  {journal}
  {Classical and Quantum Gravity}\ }\textbf {\bibinfo {volume} {29}},\ \bibinfo
  {pages} {184001} (\bibinfo {year} {2012})}\BibitemShut {NoStop}%
\bibitem [{\citenamefont {Hohensee}\ \emph {et~al.}(2013)\citenamefont
  {Hohensee}, \citenamefont {M\"uller},\ and\ \citenamefont
  {Wiringa}}]{Hohensee2013}%
  \BibitemOpen
  \bibfield  {author} {\bibinfo {author} {\bibfnamefont {M.~A.}\ \bibnamefont
  {Hohensee}}, \bibinfo {author} {\bibfnamefont {H.}~\bibnamefont {M\"uller}},
  \ and\ \bibinfo {author} {\bibfnamefont {R.~B.}\ \bibnamefont {Wiringa}},\
  }\href@noop {} {\bibfield  {journal} {\bibinfo  {journal} {Physical Review
  Letters}\ }\textbf {\bibinfo {volume} {111}},\ \bibinfo {pages} {151102}
  (\bibinfo {year} {2013})}\BibitemShut {NoStop}%
\bibitem [{\citenamefont {Dimopoulos}\ \emph {et~al.}(2007)\citenamefont
  {Dimopoulos}, \citenamefont {Graham}, \citenamefont {Hogan},\ and\
  \citenamefont {Kasevich}}]{Dimopoulos2007}%
  \BibitemOpen
  \bibfield  {author} {\bibinfo {author} {\bibfnamefont {S.}~\bibnamefont
  {Dimopoulos}}, \bibinfo {author} {\bibfnamefont {P.~W.}\ \bibnamefont
  {Graham}}, \bibinfo {author} {\bibfnamefont {J.~M.}\ \bibnamefont {Hogan}}, \
  and\ \bibinfo {author} {\bibfnamefont {M.~A.}\ \bibnamefont {Kasevich}},\
  }\href@noop {} {\bibfield  {journal} {\bibinfo  {journal} {Phys. Rev. Lett.}\
  }\textbf {\bibinfo {volume} {98}},\ \bibinfo {pages} {111102} (\bibinfo
  {year} {2007})}\BibitemShut {NoStop}%
\bibitem [{\citenamefont {Zhou}\ \emph {et~al.}(2011)\citenamefont {Zhou},
  \citenamefont {Xiong}, \citenamefont {Yang}, \citenamefont {Tang},
  \citenamefont {Peng}, \citenamefont {Hao}, \citenamefont {Li}, \citenamefont
  {Liu}, \citenamefont {Wang},\ and\ \citenamefont {Zhan}}]{Zhou2011}%
  \BibitemOpen
  \bibfield  {author} {\bibinfo {author} {\bibfnamefont {L.}~\bibnamefont
  {Zhou}}, \bibinfo {author} {\bibfnamefont {Z.~Y.}\ \bibnamefont {Xiong}},
  \bibinfo {author} {\bibfnamefont {W.}~\bibnamefont {Yang}}, \bibinfo {author}
  {\bibfnamefont {B.}~\bibnamefont {Tang}}, \bibinfo {author} {\bibfnamefont
  {W.~C.}\ \bibnamefont {Peng}}, \bibinfo {author} {\bibfnamefont
  {K.}~\bibnamefont {Hao}}, \bibinfo {author} {\bibfnamefont {R.~B.}\
  \bibnamefont {Li}}, \bibinfo {author} {\bibfnamefont {J.}~\bibnamefont
  {Liu}}, \bibinfo {author} {\bibfnamefont {J.}~\bibnamefont {Wang}}, \ and\
  \bibinfo {author} {\bibfnamefont {M.~S.}\ \bibnamefont {Zhan}},\ }\href@noop
  {} {\bibfield  {journal} {\bibinfo  {journal} {General Relativity and
  Gravitation}\ }\textbf {\bibinfo {volume} {43}},\ \bibinfo {pages} {1931}
  (\bibinfo {year} {2011})}\BibitemShut {NoStop}%
\bibitem [{\citenamefont {Hartwig}\ \emph {et~al.}(2015)\citenamefont
  {Hartwig}, \citenamefont {Abend}, \citenamefont {Schubert}, \citenamefont
  {Schlippert}, \citenamefont {Ahlers}, \citenamefont {Posso-Trujillo},
  \citenamefont {Gaaloul}, \citenamefont {Ertmer},\ and\ \citenamefont
  {Rasel}}]{Hartwig2015}%
  \BibitemOpen
  \bibfield  {author} {\bibinfo {author} {\bibfnamefont {J.}~\bibnamefont
  {Hartwig}}, \bibinfo {author} {\bibfnamefont {S.}~\bibnamefont {Abend}},
  \bibinfo {author} {\bibfnamefont {C.}~\bibnamefont {Schubert}}, \bibinfo
  {author} {\bibfnamefont {D.}~\bibnamefont {Schlippert}}, \bibinfo {author}
  {\bibfnamefont {H.}~\bibnamefont {Ahlers}}, \bibinfo {author} {\bibfnamefont
  {K.}~\bibnamefont {Posso-Trujillo}}, \bibinfo {author} {\bibfnamefont
  {N.}~\bibnamefont {Gaaloul}}, \bibinfo {author} {\bibfnamefont
  {W.}~\bibnamefont {Ertmer}}, \ and\ \bibinfo {author} {\bibfnamefont {E.~M.}\
  \bibnamefont {Rasel}},\ }\href@noop {} {\bibfield  {journal} {\bibinfo
  {journal} {arXiv}\ }\textbf {\bibinfo {volume} {1503.01213v1}} (\bibinfo
  {year} {2015})}\BibitemShut {NoStop}%
\bibitem [{\citenamefont {M\"untinga}\ \emph {et~al.}(2013)\citenamefont
  {M\"untinga}, \citenamefont {Ahlers}, \citenamefont {Krutzik}, \citenamefont
  {Wenzlawski} \emph {et~al.}}]{Muntinga2013}%
  \BibitemOpen
  \bibfield  {author} {\bibinfo {author} {\bibfnamefont {H.}~\bibnamefont
  {M\"untinga}}, \bibinfo {author} {\bibfnamefont {H.}~\bibnamefont {Ahlers}},
  \bibinfo {author} {\bibfnamefont {M.}~\bibnamefont {Krutzik}}, \bibinfo
  {author} {\bibfnamefont {A.}~\bibnamefont {Wenzlawski}},  \emph {et~al.},\
  }\href@noop {} {\bibfield  {journal} {\bibinfo  {journal} {Phys. Rev. Lett.}\
  }\textbf {\bibinfo {volume} {110}},\ \bibinfo {pages} {093602} (\bibinfo
  {year} {2013})}\BibitemShut {NoStop}%
\bibitem [{\citenamefont {Geiger}\ \emph {et~al.}(2011)\citenamefont {Geiger},
  \citenamefont {M\'{e}noret}, \citenamefont {Stern}, \citenamefont {Zahzam},
  \citenamefont {Cheinet}, \citenamefont {Battelier}, \citenamefont {Villing},
  \citenamefont {Moron}, \citenamefont {Lours}, \citenamefont {Bidel},
  \citenamefont {Bresson}, \citenamefont {Landragin},\ and\ \citenamefont
  {Bouyer}}]{Geiger2011}%
  \BibitemOpen
  \bibfield  {author} {\bibinfo {author} {\bibfnamefont {R.}~\bibnamefont
  {Geiger}}, \bibinfo {author} {\bibfnamefont {V.}~\bibnamefont {M\'{e}noret}},
  \bibinfo {author} {\bibfnamefont {G.}~\bibnamefont {Stern}}, \bibinfo
  {author} {\bibfnamefont {N.}~\bibnamefont {Zahzam}}, \bibinfo {author}
  {\bibfnamefont {P.}~\bibnamefont {Cheinet}}, \bibinfo {author} {\bibfnamefont
  {B.}~\bibnamefont {Battelier}}, \bibinfo {author} {\bibfnamefont
  {A.}~\bibnamefont {Villing}}, \bibinfo {author} {\bibfnamefont
  {F.}~\bibnamefont {Moron}}, \bibinfo {author} {\bibfnamefont
  {M.}~\bibnamefont {Lours}}, \bibinfo {author} {\bibfnamefont
  {Y.}~\bibnamefont {Bidel}}, \bibinfo {author} {\bibfnamefont
  {A.}~\bibnamefont {Bresson}}, \bibinfo {author} {\bibfnamefont
  {A.}~\bibnamefont {Landragin}}, \ and\ \bibinfo {author} {\bibfnamefont
  {P.}~\bibnamefont {Bouyer}},\ }\href@noop {} {\bibfield  {journal} {\bibinfo
  {journal} {Nature Commun.}\ }\textbf {\bibinfo {volume} {2}},\ \bibinfo
  {pages} {474} (\bibinfo {year} {2011})}\BibitemShut {NoStop}%
\bibitem [{\citenamefont {Tino}\ \emph {et~al.}(2013)\citenamefont {Tino},
  \citenamefont {Sorrentino}, \citenamefont {Aguilera}, \citenamefont
  {Battelier}, \citenamefont {Bertoldi}, \citenamefont {Bodart}, \citenamefont
  {Bongs}, \citenamefont {Bouyer}, \citenamefont {Braxmaier}, \citenamefont
  {Cacciapuoti}, \citenamefont {Gaaloul}, \citenamefont {Gürlebeck},
  \citenamefont {Hauth}, \citenamefont {Herrmann}, \citenamefont {Krutzik},
  \citenamefont {Kubelka}, \citenamefont {Landragin}, \citenamefont {Milke},
  \citenamefont {Peters}, \citenamefont {Rasel}, \citenamefont {Rocco},
  \citenamefont {Schubert}, \citenamefont {Schuldt}, \citenamefont
  {Sengstock},\ and\ \citenamefont {Wicht}}]{Tino2013}%
  \BibitemOpen
  \bibfield  {author} {\bibinfo {author} {\bibfnamefont {G.}~\bibnamefont
  {Tino}}, \bibinfo {author} {\bibfnamefont {F.}~\bibnamefont {Sorrentino}},
  \bibinfo {author} {\bibfnamefont {D.}~\bibnamefont {Aguilera}}, \bibinfo
  {author} {\bibfnamefont {B.}~\bibnamefont {Battelier}}, \bibinfo {author}
  {\bibfnamefont {A.}~\bibnamefont {Bertoldi}}, \bibinfo {author}
  {\bibfnamefont {Q.}~\bibnamefont {Bodart}}, \bibinfo {author} {\bibfnamefont
  {K.}~\bibnamefont {Bongs}}, \bibinfo {author} {\bibfnamefont
  {P.}~\bibnamefont {Bouyer}}, \bibinfo {author} {\bibfnamefont
  {C.}~\bibnamefont {Braxmaier}}, \bibinfo {author} {\bibfnamefont
  {L.}~\bibnamefont {Cacciapuoti}}, \bibinfo {author} {\bibfnamefont
  {N.}~\bibnamefont {Gaaloul}}, \bibinfo {author} {\bibfnamefont
  {N.}~\bibnamefont {Gürlebeck}}, \bibinfo {author} {\bibfnamefont
  {M.}~\bibnamefont {Hauth}}, \bibinfo {author} {\bibfnamefont
  {S.}~\bibnamefont {Herrmann}}, \bibinfo {author} {\bibfnamefont
  {M.}~\bibnamefont {Krutzik}}, \bibinfo {author} {\bibfnamefont
  {A.}~\bibnamefont {Kubelka}}, \bibinfo {author} {\bibfnamefont
  {A.}~\bibnamefont {Landragin}}, \bibinfo {author} {\bibfnamefont
  {A.}~\bibnamefont {Milke}}, \bibinfo {author} {\bibfnamefont
  {A.}~\bibnamefont {Peters}}, \bibinfo {author} {\bibfnamefont
  {E.}~\bibnamefont {Rasel}}, \bibinfo {author} {\bibfnamefont
  {E.}~\bibnamefont {Rocco}}, \bibinfo {author} {\bibfnamefont
  {C.}~\bibnamefont {Schubert}}, \bibinfo {author} {\bibfnamefont
  {T.}~\bibnamefont {Schuldt}}, \bibinfo {author} {\bibfnamefont
  {K.}~\bibnamefont {Sengstock}}, \ and\ \bibinfo {author} {\bibfnamefont
  {A.}~\bibnamefont {Wicht}},\ }\href@noop {} {\bibfield  {journal} {\bibinfo
  {journal} {Nucl. Phys. B}\ }\textbf {\bibinfo {volume} {243 - 244}},\
  \bibinfo {pages} {203 } (\bibinfo {year} {2013})}\BibitemShut {NoStop}%
\bibitem [{\citenamefont {Altschul}\ \emph {et~al.}(2015)\citenamefont
  {Altschul}, \citenamefont {Bailey}, \citenamefont {Blanchet}, \citenamefont
  {Bongs}, \citenamefont {Bouyer}, \citenamefont {Cacciapuoti}, \citenamefont
  {Capozziello}, \citenamefont {Gaaloul}, \citenamefont {Giulini},
  \citenamefont {Hartwig}, \citenamefont {Iess}, \citenamefont {Jetzer},
  \citenamefont {Landragin}, \citenamefont {Rasel}, \citenamefont {Reynaud},
  \citenamefont {Schiller}, \citenamefont {Schubert}, \citenamefont
  {Sorrentino}, \citenamefont {Sterr}, \citenamefont {Tasson}, \citenamefont
  {Tino}, \citenamefont {Tuckey},\ and\ \citenamefont {Wolf}}]{Altschul2015}%
  \BibitemOpen
  \bibfield  {author} {\bibinfo {author} {\bibfnamefont {B.}~\bibnamefont
  {Altschul}}, \bibinfo {author} {\bibfnamefont {Q.~G.}\ \bibnamefont
  {Bailey}}, \bibinfo {author} {\bibfnamefont {L.}~\bibnamefont {Blanchet}},
  \bibinfo {author} {\bibfnamefont {K.}~\bibnamefont {Bongs}}, \bibinfo
  {author} {\bibfnamefont {P.}~\bibnamefont {Bouyer}}, \bibinfo {author}
  {\bibfnamefont {L.}~\bibnamefont {Cacciapuoti}}, \bibinfo {author}
  {\bibfnamefont {S.}~\bibnamefont {Capozziello}}, \bibinfo {author}
  {\bibfnamefont {N.}~\bibnamefont {Gaaloul}}, \bibinfo {author} {\bibfnamefont
  {D.}~\bibnamefont {Giulini}}, \bibinfo {author} {\bibfnamefont
  {J.}~\bibnamefont {Hartwig}}, \bibinfo {author} {\bibfnamefont
  {L.}~\bibnamefont {Iess}}, \bibinfo {author} {\bibfnamefont {P.}~\bibnamefont
  {Jetzer}}, \bibinfo {author} {\bibfnamefont {A.}~\bibnamefont {Landragin}},
  \bibinfo {author} {\bibfnamefont {E.}~\bibnamefont {Rasel}}, \bibinfo
  {author} {\bibfnamefont {S.}~\bibnamefont {Reynaud}}, \bibinfo {author}
  {\bibfnamefont {S.}~\bibnamefont {Schiller}}, \bibinfo {author}
  {\bibfnamefont {C.}~\bibnamefont {Schubert}}, \bibinfo {author}
  {\bibfnamefont {F.}~\bibnamefont {Sorrentino}}, \bibinfo {author}
  {\bibfnamefont {U.}~\bibnamefont {Sterr}}, \bibinfo {author} {\bibfnamefont
  {J.~D.}\ \bibnamefont {Tasson}}, \bibinfo {author} {\bibfnamefont {G.~M.}\
  \bibnamefont {Tino}}, \bibinfo {author} {\bibfnamefont {P.}~\bibnamefont
  {Tuckey}}, \ and\ \bibinfo {author} {\bibfnamefont {P.}~\bibnamefont
  {Wolf}},\ }\href@noop {} {\bibfield  {journal} {\bibinfo  {journal} {Advances
  in Space Research}\ }\textbf {\bibinfo {volume} {55}},\ \bibinfo {pages} {501
  } (\bibinfo {year} {2015})}\BibitemShut {NoStop}%
\bibitem [{\citenamefont {Aguilera}\ \emph {et~al.}(2014)\citenamefont
  {Aguilera}, \citenamefont {Ahlers}, \citenamefont {Battelier}, \citenamefont
  {Bawamia}, \citenamefont {Bertoldi} \emph {et~al.}}]{Aguilera2014}%
  \BibitemOpen
  \bibfield  {author} {\bibinfo {author} {\bibfnamefont {D.}~\bibnamefont
  {Aguilera}}, \bibinfo {author} {\bibfnamefont {H.}~\bibnamefont {Ahlers}},
  \bibinfo {author} {\bibfnamefont {B.}~\bibnamefont {Battelier}}, \bibinfo
  {author} {\bibfnamefont {A.}~\bibnamefont {Bawamia}}, \bibinfo {author}
  {\bibfnamefont {A.}~\bibnamefont {Bertoldi}},  \emph {et~al.},\ }\href@noop
  {} {\bibfield  {journal} {\bibinfo  {journal} {Class.Quant.Grav.}\ }\textbf
  {\bibinfo {volume} {31}},\ \bibinfo {pages} {115010} (\bibinfo {year}
  {2014})}\BibitemShut {NoStop}%
\bibitem [{\citenamefont {L\'ev\`eque}\ \emph {et~al.}(2009)\citenamefont
  {L\'ev\`eque}, \citenamefont {Gauguet}, \citenamefont {Michaud},
  \citenamefont {Pereira Dos~Santos},\ and\ \citenamefont
  {Landragin}}]{Leveque2009}%
  \BibitemOpen
  \bibfield  {author} {\bibinfo {author} {\bibfnamefont {T.}~\bibnamefont
  {L\'ev\`eque}}, \bibinfo {author} {\bibfnamefont {A.}~\bibnamefont
  {Gauguet}}, \bibinfo {author} {\bibfnamefont {F.}~\bibnamefont {Michaud}},
  \bibinfo {author} {\bibfnamefont {F.}~\bibnamefont {Pereira Dos~Santos}}, \
  and\ \bibinfo {author} {\bibfnamefont {A.}~\bibnamefont {Landragin}},\
  }\href@noop {} {\bibfield  {journal} {\bibinfo  {journal} {Phys. Rev. Lett.}\
  }\textbf {\bibinfo {volume} {103}},\ \bibinfo {pages} {080405} (\bibinfo
  {year} {2009})}\BibitemShut {NoStop}%
\bibitem [{\citenamefont {Chiow}\ \emph {et~al.}(2011)\citenamefont {Chiow},
  \citenamefont {Kovachy}, \citenamefont {Chien},\ and\ \citenamefont
  {Kasevich}}]{Chiow2011}%
  \BibitemOpen
  \bibfield  {author} {\bibinfo {author} {\bibfnamefont {S.-w.}\ \bibnamefont
  {Chiow}}, \bibinfo {author} {\bibfnamefont {T.}~\bibnamefont {Kovachy}},
  \bibinfo {author} {\bibfnamefont {H.-C.}\ \bibnamefont {Chien}}, \ and\
  \bibinfo {author} {\bibfnamefont {M.~A.}\ \bibnamefont {Kasevich}},\
  }\href@noop {} {\bibfield  {journal} {\bibinfo  {journal} {Phys. Rev. Lett.}\
  }\textbf {\bibinfo {volume} {107}},\ \bibinfo {pages} {130403} (\bibinfo
  {year} {2011})}\BibitemShut {NoStop}%
\bibitem [{\citenamefont {Clad\'e}\ \emph {et~al.}(2009)\citenamefont
  {Clad\'e}, \citenamefont {Guellati-Kh\'elifa}, \citenamefont {Nez},\ and\
  \citenamefont {Biraben}}]{Clade2009}%
  \BibitemOpen
  \bibfield  {author} {\bibinfo {author} {\bibfnamefont {P.}~\bibnamefont
  {Clad\'e}}, \bibinfo {author} {\bibfnamefont {S.}~\bibnamefont
  {Guellati-Kh\'elifa}}, \bibinfo {author} {\bibfnamefont {F.}~\bibnamefont
  {Nez}}, \ and\ \bibinfo {author} {\bibfnamefont {F.}~\bibnamefont
  {Biraben}},\ }\href@noop {} {\bibfield  {journal} {\bibinfo  {journal}
  {Physical Review Letters}\ }\textbf {\bibinfo {volume} {102}},\ \bibinfo
  {pages} {240402} (\bibinfo {year} {2009})}\BibitemShut {NoStop}%
\bibitem [{\citenamefont {M\"uller}\ \emph {et~al.}(2009)\citenamefont
  {M\"uller}, \citenamefont {Chiow}, \citenamefont {Herrmann},\ and\
  \citenamefont {Chu}}]{Muller2009}%
  \BibitemOpen
  \bibfield  {author} {\bibinfo {author} {\bibfnamefont {H.}~\bibnamefont
  {M\"uller}}, \bibinfo {author} {\bibfnamefont {S.-W.}\ \bibnamefont {Chiow}},
  \bibinfo {author} {\bibfnamefont {S.}~\bibnamefont {Herrmann}}, \ and\
  \bibinfo {author} {\bibfnamefont {S.}~\bibnamefont {Chu}},\ }\href@noop {}
  {\bibfield  {journal} {\bibinfo  {journal} {Physical Review Letters}\
  }\textbf {\bibinfo {volume} {102}},\ \bibinfo {pages} {240403} (\bibinfo
  {year} {2009})}\BibitemShut {NoStop}%
\bibitem [{\citenamefont {Hardman}\ \emph {et~al.}(2014)\citenamefont
  {Hardman}, \citenamefont {Kuhn}, \citenamefont {McDonald}, \citenamefont
  {Debs}, \citenamefont {Bennetts}, \citenamefont {Close},\ and\ \citenamefont
  {Robins}}]{Hardman2014}%
  \BibitemOpen
  \bibfield  {author} {\bibinfo {author} {\bibfnamefont {K.~S.}\ \bibnamefont
  {Hardman}}, \bibinfo {author} {\bibfnamefont {C.~C.~N.}\ \bibnamefont
  {Kuhn}}, \bibinfo {author} {\bibfnamefont {G.~D.}\ \bibnamefont {McDonald}},
  \bibinfo {author} {\bibfnamefont {J.~E.}\ \bibnamefont {Debs}}, \bibinfo
  {author} {\bibfnamefont {S.}~\bibnamefont {Bennetts}}, \bibinfo {author}
  {\bibfnamefont {J.~D.}\ \bibnamefont {Close}}, \ and\ \bibinfo {author}
  {\bibfnamefont {N.~P.}\ \bibnamefont {Robins}},\ }\href@noop {} {\bibfield
  {journal} {\bibinfo  {journal} {Phys. Rev. A}\ }\textbf {\bibinfo {volume}
  {89}},\ \bibinfo {pages} {023626} (\bibinfo {year} {2014})}\BibitemShut
  {NoStop}%
\bibitem [{\citenamefont {Kuhn}\ \emph {et~al.}(2014)\citenamefont {Kuhn},
  \citenamefont {McDonald}, \citenamefont {Hardman}, \citenamefont {Bennetts},
  \citenamefont {Everitt}, \citenamefont {Altin}, \citenamefont {Debs},
  \citenamefont {Close},\ and\ \citenamefont {Robins}}]{Kuhn2014}%
  \BibitemOpen
  \bibfield  {author} {\bibinfo {author} {\bibfnamefont {C.~C.~N.}\
  \bibnamefont {Kuhn}}, \bibinfo {author} {\bibfnamefont {G.~D.}\ \bibnamefont
  {McDonald}}, \bibinfo {author} {\bibfnamefont {K.~S.}\ \bibnamefont
  {Hardman}}, \bibinfo {author} {\bibfnamefont {S.}~\bibnamefont {Bennetts}},
  \bibinfo {author} {\bibfnamefont {P.~J.}\ \bibnamefont {Everitt}}, \bibinfo
  {author} {\bibfnamefont {P.~A.}\ \bibnamefont {Altin}}, \bibinfo {author}
  {\bibfnamefont {J.~E.}\ \bibnamefont {Debs}}, \bibinfo {author}
  {\bibfnamefont {J.~D.}\ \bibnamefont {Close}}, \ and\ \bibinfo {author}
  {\bibfnamefont {N.~P.}\ \bibnamefont {Robins}},\ }\href@noop {} {\bibfield
  {journal} {\bibinfo  {journal} {New J. Phys.}\ }\textbf {\bibinfo {volume}
  {16}},\ \bibinfo {pages} {073035} (\bibinfo {year} {2014})}\BibitemShut
  {NoStop}%
\bibitem [{\citenamefont {M\"uller}\ \emph {et~al.}(2008)\citenamefont
  {M\"uller}, \citenamefont {w.~Chiow}, \citenamefont {Herrmann}, \citenamefont
  {Chu},\ and\ \citenamefont {Chung}}]{Muller2008}%
  \BibitemOpen
  \bibfield  {author} {\bibinfo {author} {\bibfnamefont {H.}~\bibnamefont
  {M\"uller}}, \bibinfo {author} {\bibfnamefont {S.}~\bibnamefont {w.~Chiow}},
  \bibinfo {author} {\bibfnamefont {S.}~\bibnamefont {Herrmann}}, \bibinfo
  {author} {\bibfnamefont {S.}~\bibnamefont {Chu}}, \ and\ \bibinfo {author}
  {\bibfnamefont {K.-Y.}\ \bibnamefont {Chung}},\ }\href@noop {} {\bibfield
  {journal} {\bibinfo  {journal} {Phys. Rev. Lett.}\ }\textbf {\bibinfo
  {volume} {100}},\ \bibinfo {pages} {031101} (\bibinfo {year}
  {2008})}\BibitemShut {NoStop}%
\bibitem [{\citenamefont {Kasevich}\ and\ \citenamefont
  {Chu}(1992)}]{Kasevich1992}%
  \BibitemOpen
  \bibfield  {author} {\bibinfo {author} {\bibfnamefont {M.}~\bibnamefont
  {Kasevich}}\ and\ \bibinfo {author} {\bibfnamefont {S.}~\bibnamefont {Chu}},\
  }\href@noop {} {\bibfield  {journal} {\bibinfo  {journal} {Appl. Phys. B}\
  }\textbf {\bibinfo {volume} {54}},\ \bibinfo {pages} {321} (\bibinfo {year}
  {1992})}\BibitemShut {NoStop}%
\bibitem [{\citenamefont {Cheinet}\ \emph {et~al.}(2008)\citenamefont
  {Cheinet}, \citenamefont {Canuel}, \citenamefont {Pereira Dos~Santos},
  \citenamefont {Gauguet}, \citenamefont {Leduc},\ and\ \citenamefont
  {Landragin}}]{Cheinet2008}%
  \BibitemOpen
  \bibfield  {author} {\bibinfo {author} {\bibfnamefont {P.}~\bibnamefont
  {Cheinet}}, \bibinfo {author} {\bibfnamefont {B.}~\bibnamefont {Canuel}},
  \bibinfo {author} {\bibfnamefont {F.}~\bibnamefont {Pereira Dos~Santos}},
  \bibinfo {author} {\bibfnamefont {A.}~\bibnamefont {Gauguet}}, \bibinfo
  {author} {\bibfnamefont {F.}~\bibnamefont {Leduc}}, \ and\ \bibinfo {author}
  {\bibfnamefont {A.}~\bibnamefont {Landragin}},\ }\href@noop {} {\bibfield
  {journal} {\bibinfo  {journal} {IEEE Trans. Instrum. Meas.}\ }\textbf
  {\bibinfo {volume} {57}},\ \bibinfo {pages} {1141} (\bibinfo {year}
  {2008})}\BibitemShut {NoStop}%
\bibitem [{\citenamefont {Foster}\ \emph {et~al.}(2002)\citenamefont {Foster},
  \citenamefont {Fixler}, \citenamefont {McGuirk},\ and\ \citenamefont
  {Kasevich}}]{Foster2002}%
  \BibitemOpen
  \bibfield  {author} {\bibinfo {author} {\bibfnamefont {G.~T.}\ \bibnamefont
  {Foster}}, \bibinfo {author} {\bibfnamefont {J.~B.}\ \bibnamefont {Fixler}},
  \bibinfo {author} {\bibfnamefont {J.~M.}\ \bibnamefont {McGuirk}}, \ and\
  \bibinfo {author} {\bibfnamefont {M.~A.}\ \bibnamefont {Kasevich}},\
  }\href@noop {} {\bibfield  {journal} {\bibinfo  {journal} {Optics Letters}\
  }\textbf {\bibinfo {volume} {27}},\ \bibinfo {pages} {951} (\bibinfo {year}
  {2002})}\BibitemShut {NoStop}%
\bibitem [{\citenamefont {Stockton}\ \emph {et~al.}(2007)\citenamefont
  {Stockton}, \citenamefont {Wu},\ and\ \citenamefont
  {Kasevich}}]{Stockton2007}%
  \BibitemOpen
  \bibfield  {author} {\bibinfo {author} {\bibfnamefont {J.~K.}\ \bibnamefont
  {Stockton}}, \bibinfo {author} {\bibfnamefont {X.}~\bibnamefont {Wu}}, \ and\
  \bibinfo {author} {\bibfnamefont {M.~A.}\ \bibnamefont {Kasevich}},\
  }\href@noop {} {\bibfield  {journal} {\bibinfo  {journal} {Phys. Rev. A}\
  }\textbf {\bibinfo {volume} {76}},\ \bibinfo {pages} {033613} (\bibinfo
  {year} {2007})}\BibitemShut {NoStop}%
\bibitem [{\citenamefont {Varoquaux}\ \emph {et~al.}(2009)\citenamefont
  {Varoquaux}, \citenamefont {Nyman}, \citenamefont {Geiger}, \citenamefont
  {Cheinet}, \citenamefont {Landragin},\ and\ \citenamefont
  {Bouyer}}]{Varoquaux2009}%
  \BibitemOpen
  \bibfield  {author} {\bibinfo {author} {\bibfnamefont {G.}~\bibnamefont
  {Varoquaux}}, \bibinfo {author} {\bibfnamefont {R.~A.}\ \bibnamefont
  {Nyman}}, \bibinfo {author} {\bibfnamefont {R.}~\bibnamefont {Geiger}},
  \bibinfo {author} {\bibfnamefont {P.}~\bibnamefont {Cheinet}}, \bibinfo
  {author} {\bibfnamefont {A.}~\bibnamefont {Landragin}}, \ and\ \bibinfo
  {author} {\bibfnamefont {P.}~\bibnamefont {Bouyer}},\ }\href@noop {}
  {\bibfield  {journal} {\bibinfo  {journal} {New Journal of Physics}\ }\textbf
  {\bibinfo {volume} {11}},\ \bibinfo {pages} {113010} (\bibinfo {year}
  {2009})}\BibitemShut {NoStop}%
\bibitem [{\citenamefont {Chen}\ \emph {et~al.}(2014)\citenamefont {Chen},
  \citenamefont {Zhong}, \citenamefont {Song}, \citenamefont {Zhu},
  \citenamefont {Wang},\ and\ \citenamefont {Zhan}}]{Chen2014}%
  \BibitemOpen
  \bibfield  {author} {\bibinfo {author} {\bibfnamefont {X.}~\bibnamefont
  {Chen}}, \bibinfo {author} {\bibfnamefont {J.}~\bibnamefont {Zhong}},
  \bibinfo {author} {\bibfnamefont {H.}~\bibnamefont {Song}}, \bibinfo {author}
  {\bibfnamefont {L.}~\bibnamefont {Zhu}}, \bibinfo {author} {\bibfnamefont
  {J.}~\bibnamefont {Wang}}, \ and\ \bibinfo {author} {\bibfnamefont
  {M.}~\bibnamefont {Zhan}},\ }\href@noop {} {\bibfield  {journal} {\bibinfo
  {journal} {Phys. Rev. A}\ }\textbf {\bibinfo {volume} {90}},\ \bibinfo
  {pages} {023609} (\bibinfo {year} {2014})}\BibitemShut {NoStop}%
\bibitem [{\citenamefont {Wu}(2009)}]{Wu_thesis}%
  \BibitemOpen
  \bibfield  {author} {\bibinfo {author} {\bibfnamefont {X.}~\bibnamefont
  {Wu}},\ }\emph {\bibinfo {title} {Gravity gradient survey with a mobile atom
  interferometer}},\ \href@noop {} {\bibinfo {type} {{Ph.D} thesis}},\ \bibinfo
   {school} {Stanford University} (\bibinfo {year} {2009}),\ \bibinfo {note}
  {{C}hapter 5}\BibitemShut {NoStop}%
\bibitem [{\citenamefont {Carraz}\ \emph {et~al.}(2009)\citenamefont {Carraz},
  \citenamefont {Lienhart}, \citenamefont {Charri\`{e}re}, \citenamefont
  {Cadoret}, \citenamefont {Zahzam}, \citenamefont {Bidel},\ and\ \citenamefont
  {Bresson}}]{Carraz2009}%
  \BibitemOpen
  \bibfield  {author} {\bibinfo {author} {\bibfnamefont {O.}~\bibnamefont
  {Carraz}}, \bibinfo {author} {\bibfnamefont {F.}~\bibnamefont {Lienhart}},
  \bibinfo {author} {\bibfnamefont {R.}~\bibnamefont {Charri\`{e}re}}, \bibinfo
  {author} {\bibfnamefont {M.}~\bibnamefont {Cadoret}}, \bibinfo {author}
  {\bibfnamefont {N.}~\bibnamefont {Zahzam}}, \bibinfo {author} {\bibfnamefont
  {Y.}~\bibnamefont {Bidel}}, \ and\ \bibinfo {author} {\bibfnamefont
  {A.}~\bibnamefont {Bresson}},\ }\href@noop {} {\bibfield  {journal} {\bibinfo
   {journal} {Appl. Phys. B}\ }\textbf {\bibinfo {volume} {97}},\ \bibinfo
  {pages} {405} (\bibinfo {year} {2009})}\BibitemShut {NoStop}%
\bibitem [{\citenamefont {Carraz}\ \emph {et~al.}(2012)\citenamefont {Carraz},
  \citenamefont {Charri\`{e}re}, \citenamefont {Cadoret}, \citenamefont
  {Zahzam}, \citenamefont {Bidel},\ and\ \citenamefont {Bresson}}]{Carraz2012}%
  \BibitemOpen
  \bibfield  {author} {\bibinfo {author} {\bibfnamefont {O.}~\bibnamefont
  {Carraz}}, \bibinfo {author} {\bibfnamefont {R.}~\bibnamefont
  {Charri\`{e}re}}, \bibinfo {author} {\bibfnamefont {M.}~\bibnamefont
  {Cadoret}}, \bibinfo {author} {\bibfnamefont {N.}~\bibnamefont {Zahzam}},
  \bibinfo {author} {\bibfnamefont {Y.}~\bibnamefont {Bidel}}, \ and\ \bibinfo
  {author} {\bibfnamefont {A.}~\bibnamefont {Bresson}},\ }\href@noop {}
  {\bibfield  {journal} {\bibinfo  {journal} {Phys. Rev. A}\ }\textbf {\bibinfo
  {volume} {86}},\ \bibinfo {pages} {033605} (\bibinfo {year}
  {2012})}\BibitemShut {NoStop}%
\bibitem [{\citenamefont {S\"uptitz}\ \emph {et~al.}(1994)\citenamefont
  {S\"uptitz}, \citenamefont {Wokurka}, \citenamefont {Strauch}, \citenamefont
  {Kohns},\ and\ \citenamefont {Ertmer}}]{Suptitz1994}%
  \BibitemOpen
  \bibfield  {author} {\bibinfo {author} {\bibfnamefont {W.}~\bibnamefont
  {S\"uptitz}}, \bibinfo {author} {\bibfnamefont {G.}~\bibnamefont {Wokurka}},
  \bibinfo {author} {\bibfnamefont {F.}~\bibnamefont {Strauch}}, \bibinfo
  {author} {\bibfnamefont {P.}~\bibnamefont {Kohns}}, \ and\ \bibinfo {author}
  {\bibfnamefont {W.}~\bibnamefont {Ertmer}},\ }\href@noop {} {\bibfield
  {journal} {\bibinfo  {journal} {Opt. Lett.}\ }\textbf {\bibinfo {volume}
  {19}},\ \bibinfo {pages} {1571} (\bibinfo {year} {1994})}\BibitemShut
  {NoStop}%
\bibitem [{\citenamefont {Kasevich}\ \emph {et~al.}(1991)\citenamefont
  {Kasevich}, \citenamefont {Weiss}, \citenamefont {Riis}, \citenamefont
  {Moler}, \citenamefont {Kasapi},\ and\ \citenamefont {Chu}}]{Kasevich1991}%
  \BibitemOpen
  \bibfield  {author} {\bibinfo {author} {\bibfnamefont {M.}~\bibnamefont
  {Kasevich}}, \bibinfo {author} {\bibfnamefont {D.~S.}\ \bibnamefont {Weiss}},
  \bibinfo {author} {\bibfnamefont {E.}~\bibnamefont {Riis}}, \bibinfo {author}
  {\bibfnamefont {K.}~\bibnamefont {Moler}}, \bibinfo {author} {\bibfnamefont
  {S.}~\bibnamefont {Kasapi}}, \ and\ \bibinfo {author} {\bibfnamefont
  {S.}~\bibnamefont {Chu}},\ }\href@noop {} {\bibfield  {journal} {\bibinfo
  {journal} {Phys. Rev. Lett.}\ }\textbf {\bibinfo {volume} {66}},\ \bibinfo
  {pages} {2297} (\bibinfo {year} {1991})}\BibitemShut {NoStop}%
\bibitem [{\citenamefont {Lan}\ \emph {et~al.}(2012)\citenamefont {Lan},
  \citenamefont {Kuan}, \citenamefont {Estey}, \citenamefont {Haslinger},\ and\
  \citenamefont {M\"uller}}]{Lan2012}%
  \BibitemOpen
  \bibfield  {author} {\bibinfo {author} {\bibfnamefont {S.-Y.}\ \bibnamefont
  {Lan}}, \bibinfo {author} {\bibfnamefont {P.-C.}\ \bibnamefont {Kuan}},
  \bibinfo {author} {\bibfnamefont {B.}~\bibnamefont {Estey}}, \bibinfo
  {author} {\bibfnamefont {P.}~\bibnamefont {Haslinger}}, \ and\ \bibinfo
  {author} {\bibfnamefont {H.}~\bibnamefont {M\"uller}},\ }\href@noop {}
  {\bibfield  {journal} {\bibinfo  {journal} {Phys. Rev. Lett.}\ }\textbf
  {\bibinfo {volume} {108}},\ \bibinfo {pages} {090402} (\bibinfo {year}
  {2012})}\BibitemShut {NoStop}%
\bibitem [{\citenamefont {Louchet-Chauvin}\ \emph {et~al.}(2011)\citenamefont
  {Louchet-Chauvin}, \citenamefont {Farah}, \citenamefont {Bodart},
  \citenamefont {Clairon}, \citenamefont {Landragin}, \citenamefont {Merlet},\
  and\ \citenamefont {Pereira Dos~Santos}}]{Louchet2011}%
  \BibitemOpen
  \bibfield  {author} {\bibinfo {author} {\bibfnamefont {A.}~\bibnamefont
  {Louchet-Chauvin}}, \bibinfo {author} {\bibfnamefont {T.}~\bibnamefont
  {Farah}}, \bibinfo {author} {\bibfnamefont {Q.}~\bibnamefont {Bodart}},
  \bibinfo {author} {\bibfnamefont {A.}~\bibnamefont {Clairon}}, \bibinfo
  {author} {\bibfnamefont {A.}~\bibnamefont {Landragin}}, \bibinfo {author}
  {\bibfnamefont {S.}~\bibnamefont {Merlet}}, \ and\ \bibinfo {author}
  {\bibfnamefont {F.}~\bibnamefont {Pereira Dos~Santos}},\ }\href@noop {}
  {\bibfield  {journal} {\bibinfo  {journal} {New Journal of Physics}\ }\textbf
  {\bibinfo {volume} {13}},\ \bibinfo {pages} {065025} (\bibinfo {year}
  {2011})}\BibitemShut {NoStop}%
\bibitem [{\citenamefont {Yver-Leduc}\ \emph {et~al.}(2003)\citenamefont
  {Yver-Leduc}, \citenamefont {Cheinet}, \citenamefont {Fils}, \citenamefont
  {Clairon}, \citenamefont {Dimarcq}, \citenamefont {Holleville}, \citenamefont
  {Bouyer},\ and\ \citenamefont {Landragin}}]{Yver-Leduc2003}%
  \BibitemOpen
  \bibfield  {author} {\bibinfo {author} {\bibfnamefont {F.}~\bibnamefont
  {Yver-Leduc}}, \bibinfo {author} {\bibfnamefont {P.}~\bibnamefont {Cheinet}},
  \bibinfo {author} {\bibfnamefont {J.}~\bibnamefont {Fils}}, \bibinfo {author}
  {\bibfnamefont {A.}~\bibnamefont {Clairon}}, \bibinfo {author} {\bibfnamefont
  {N.}~\bibnamefont {Dimarcq}}, \bibinfo {author} {\bibfnamefont
  {D.}~\bibnamefont {Holleville}}, \bibinfo {author} {\bibfnamefont
  {P.}~\bibnamefont {Bouyer}}, \ and\ \bibinfo {author} {\bibfnamefont
  {A.}~\bibnamefont {Landragin}},\ }\href@noop {} {\bibfield  {journal}
  {\bibinfo  {journal} {J. Opt. B}\ }\textbf {\bibinfo {volume} {5}},\ \bibinfo
  {pages} {S136} (\bibinfo {year} {2003})}\BibitemShut {NoStop}%
\bibitem [{\citenamefont {D\"oring}\ \emph {et~al.}(2010)\citenamefont
  {D\"oring}, \citenamefont {McDonald}, \citenamefont {Debs}, \citenamefont
  {Filg}, \citenamefont {Altin}, \citenamefont {Bachor}, \citenamefont
  {Robins},\ and\ \citenamefont {Close}}]{Doring2010}%
  \BibitemOpen
  \bibfield  {author} {\bibinfo {author} {\bibfnamefont {D.}~\bibnamefont
  {D\"oring}}, \bibinfo {author} {\bibfnamefont {G.}~\bibnamefont {McDonald}},
  \bibinfo {author} {\bibfnamefont {J.~E.}\ \bibnamefont {Debs}}, \bibinfo
  {author} {\bibfnamefont {C.}~\bibnamefont {Filg}}, \bibinfo {author}
  {\bibfnamefont {P.~A.}\ \bibnamefont {Altin}}, \bibinfo {author}
  {\bibfnamefont {H.-A.}\ \bibnamefont {Bachor}}, \bibinfo {author}
  {\bibfnamefont {N.~P.}\ \bibnamefont {Robins}}, \ and\ \bibinfo {author}
  {\bibfnamefont {J.~D.}\ \bibnamefont {Close}},\ }\href@noop {} {\bibfield
  {journal} {\bibinfo  {journal} {Physical Review A}\ }\textbf {\bibinfo
  {volume} {81}},\ \bibinfo {pages} {043633} (\bibinfo {year}
  {2010})}\BibitemShut {NoStop}%
\bibitem [{\citenamefont {Dick}(1987)}]{Dick1987}%
  \BibitemOpen
  \bibfield  {author} {\bibinfo {author} {\bibfnamefont {G.}~\bibnamefont
  {Dick}},\ }\href@noop {} {\bibfield  {journal} {\bibinfo  {journal}
  {Proceedings of the Nineteenth Annual Precise Time and Time Interval
  Applications and Planning Meeting, Redondo Beach, CA}\ ,\ \bibinfo {pages}
  {133}} (\bibinfo {year} {1987})}\BibitemShut {NoStop}%
\end{thebibliography}%

\end{document}